\pdfoutput=1
\documentclass[aps,pre,reprint,showpacs,superscriptaddress,eqsecnum]{revtex4-1}

\usepackage{amsmath,graphicx,amssymb}
\usepackage{dcolumn}
\usepackage{bm}
\usepackage[squaren]{SIunits}
\usepackage[hypertexnames=false,colorlinks=true,linkcolor=blue,citecolor=blue]{hyperref}

\begin{document}

\title{Retarding the growth of the {Rosensweig} instability unveils a new scaling regime}
\author{Adrian Lange}
\email{Adrian.Lange@tu-dresden.de}
\affiliation{TU Dresden, Institute of Fluid Mechanics, Chair of Magnetofluiddynamics, Measuring and Automation Technology, 01062 Dresden, Germany}
\author{Christian Gollwitzer}
\email{Christian.Gollwitzer@ptb.de}
\affiliation{Experimentalphysik V, Universit\"at Bayreuth, D-95440 Bayreuth, Germany}
\author{Robin Maretzki}
\affiliation{Experimentalphysik V, Universit\"at Bayreuth, D-95440 Bayreuth, Germany}
\author{Ingo Rehberg}
\email{Ingo.Rehberg@uni-bayreuth.de}
\affiliation{Experimentalphysik V, Universit\"at Bayreuth, D-95440 Bayreuth, Germany}
\author{Reinhard Richter}
\email{Reinhard.Richter@uni-bayreuth.de}
\affiliation{Experimentalphysik V, Universit\"at Bayreuth, D-95440 Bayreuth, Germany}

\date{\today}

\begin{abstract}
Using a highly viscous magnetic fluid, the dynamics in the aftermath of the Rosensweig instability can be slowed down by more than 2000 times. In this way we expand the regime where the growth rate is predicted to scale linearly with the bifurcation parameter by six orders of magnitude, while this regime is tiny for standard ferrofluids and can not be resolved experimentally there.
We measure the growth of the pattern by means of a two-dimensional imaging technique,
and find that the slopes of the growth and decay rates  are  not the same - a qualitative discrepancy to the theoretical predictions. We solve this discrepancy by taking into account a viscosity which is assumed to be different for the growth and decay.
This may be a consequence of the measured shear thinning of the ferrofluid.
\end{abstract}


\maketitle

\section{\label{sec:intro}Introduction}
 The "pitch-drop-experiment" \cite{edgeworth84}, which received the Ig-Nobel Price in physics 2005, has brought to the attention that a fast process like drop formation \cite{eggers97,rothert01} can be retarded considerably if instead of a standard liquid like water -- it has a viscosity of $\unit{10^{-3}}{Pa\,s}$ at $\unit{20}{\,^\circ}$C -- a material like pitch, with a viscosity around $\unit{10^8}{Pa\,s}$ is selected. The funnel was filled in 1930 \cite{queensland2013}, and today "Finally the ninth Pitch Drop has fallen from the world's longest running lab experiment" \cite{thetenth} and the 10th is awaited within the next 14 years. Here the question arises whether those high viscosities may give access to so far not resolved phenomena.

\begin{figure}[b]
\begin{center}
  \includegraphics[width=8.55cm]{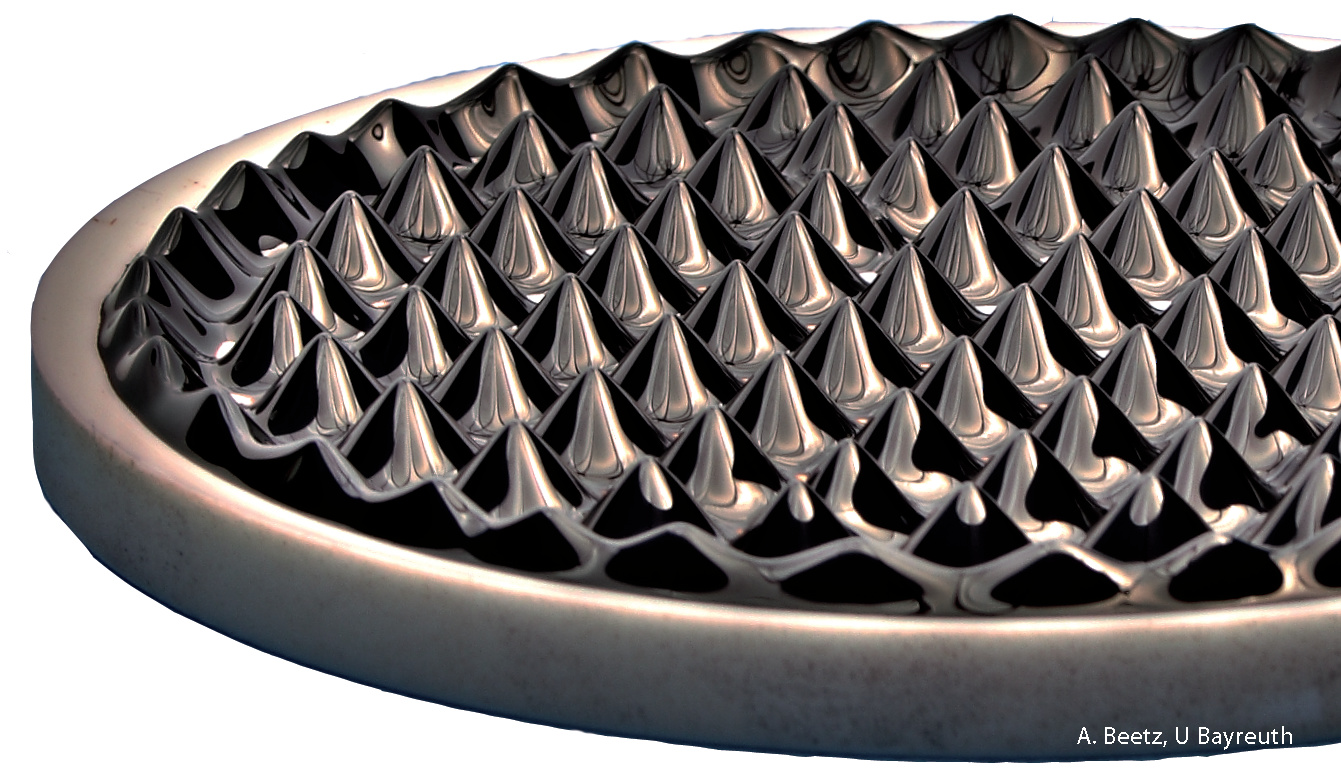}       
  \caption{Rosensweig peaks of the magnetic fluid EMG 909 at a
  supercritical induction $B > B_\mathrm{c}$ in a vessel with diameter of
  120\,mm. The picture is taken from \cite{richter11}.
  A movie showing the formation of Rosensweig patterns can be accessed at~\protect\cite{castellvecchi05}.}
  \label{fig:peaks}
\end{center}
\end{figure}

In the following we are investigating this question for the case of the well known Rosensweig or normal field instability \cite{cowley67}. It is observed in a horizontal layer of magnetic fluid (MF) \cite{rosensweig85_book}
with a free surface, when a critical value $B_\mathrm{c}$ of the vertically
oriented magnetic induction is surpassed. Figure~\ref{fig:peaks} presents a photo
of the final hexagonal arrangement of static liquid peaks. Beside the threshold, beyond which
the instability occurs, two quantities characterizing the emerging pattern have
been in the focus of various studies:
the critical wave number of the peaks and the corresponding growth rate, where
both are strongly influenced by the viscosity of the magnetic fluid.

That essential role of the viscosity for the dynamics of the pattern formation
is reflected in the course of the analyses devoted to the Rosensweig instability.
For an inviscid magnetic fluid (the dynamic viscosity $\eta$
is zero) and an infinitely deep container, \citet{cowley67} provide a linear stability
analysis already in the very first description of the normal field instability to find
the critical threshold $B_\mathrm{c}$ and the critical wave number $k_\mathrm{c}$.
This approach has been extended later by \citet{salin93}
to fluids with non-zero viscosity, where the growth rate depends on $\eta$,
and to a finite depth of the container by \citet{weilepp96}. First experimental
investigations on the growth of the pattern are provided by \citet{lange00_wave,lange01_wave},
who also derive the growth rate for the case of a viscous magnetic
fluid and an arbitrary layer thickness $h$. This theoretical analysis has been later extended
to the case of a nonlinear magnetization curve $M(H)$ by \citet{knieling07}.

Whereas so far the growth rate of the emerging Rosensweig pattern has been measured utilizing ferrofluids with
$\eta = \unit{4.2 \times 10^{-3}}{Pa\,s}$ \cite{lange00_wave,knieling07} and $\unit{5.2 \times 10^{-2}}{Pa\,s}$  \cite{knieling07} we are tackling here the growth process in a ferrofluid which is a thousand times more viscous than the first one. Such a ferrofluid is being created by cooling a commercially available viscous ferrofluid (APG E32 from Ferrotec Co.\,) down to $\unit{10}{\,^\circ}$C. The ferrofluid has now a viscosity of $\unit{(4.48 \pm 0.1)}{Pa\,s}$.
In such a cooled Rosensweig (sloppy \emph{Frozensweig}) instability \cite{gollwitzer2009frozensweig}
the growth of the pattern takes 60 seconds and can be measured
with high temporal resolution in the extended system using a two-dimensional X-ray imaging
technique \cite{richter01,gollwitzer07}. That technique provides the full surface
topography, as opposed to the \unit{7}{kHz} fast, but one dimensional Hall-sensor array, which had to be utilized for the low viscosity ferrofluids \cite{knieling07}. The potential of the retarded instability was demonstrated before \cite{gollwitzer10}, when the coefficients of nonlinear amplitude equations were determined in this way. In addition a sequence of localized patches of Rosensweig pattern could be uncovered most recently \cite{lloyd2015} with that technique.

Here we exploit a higher viscosity to investigate the linear growth rate in a regime, which was hitherto not accessible. This expectation is based on a scaling analysis presented in Ref.\,\cite{lange01_growth}. For supercritical inductions larger than $\bar\nu^2/6$ (the dimensionless kinematic viscosity $\bar\nu$ is defined in Eq.\ (\ref{eq:scaling_visc}) below) the behavior of the growth rate is characterized by a {\it square-root} dependence on those inductions, as confirmed in \cite{knieling07}. Contrary, for supercritical inductions smaller than $\bar\nu^2/6$ the behavior of the growth rate is characterized by a {\it linear} dependence.
In the present experiment we increase $\bar\nu^2/6$ by six orders of magnitude
due to the high viscosity of the ferrofluid APG E32 at $\unit{10}{\,^\circ}$C.
Thus a new territory of linear scaling is open for exploration.

The outline of the paper is as follows: the experimental setup
and the measurements are sketched in the next Sect.\ \ref{sec:experiment}.
The theoretical analysis is presented in Sect.\ \ref{sec:theory}
and compared with the experimental findings in the
subsequent Sect.\ \ref{sec:results}.

\section{\label{sec:experiment}Experiment}
In this section we describe the experimental setup (Sect.\,\ref{subsec:setup}), the ferrofluid (Sect.\,\ref{subsec:fluid}), the protocol utilized for the measurements (Sect.\,\ref{subsec:measurements}), and the way the linear growth rate is extracted from the recorded data (Sect.\,\ref{subsec:extracting}).

\subsection{\label{subsec:setup}Experimental setup}
\begin{figure*}[htbp]
\begin{center}
  \includegraphics[width=5.0cm]{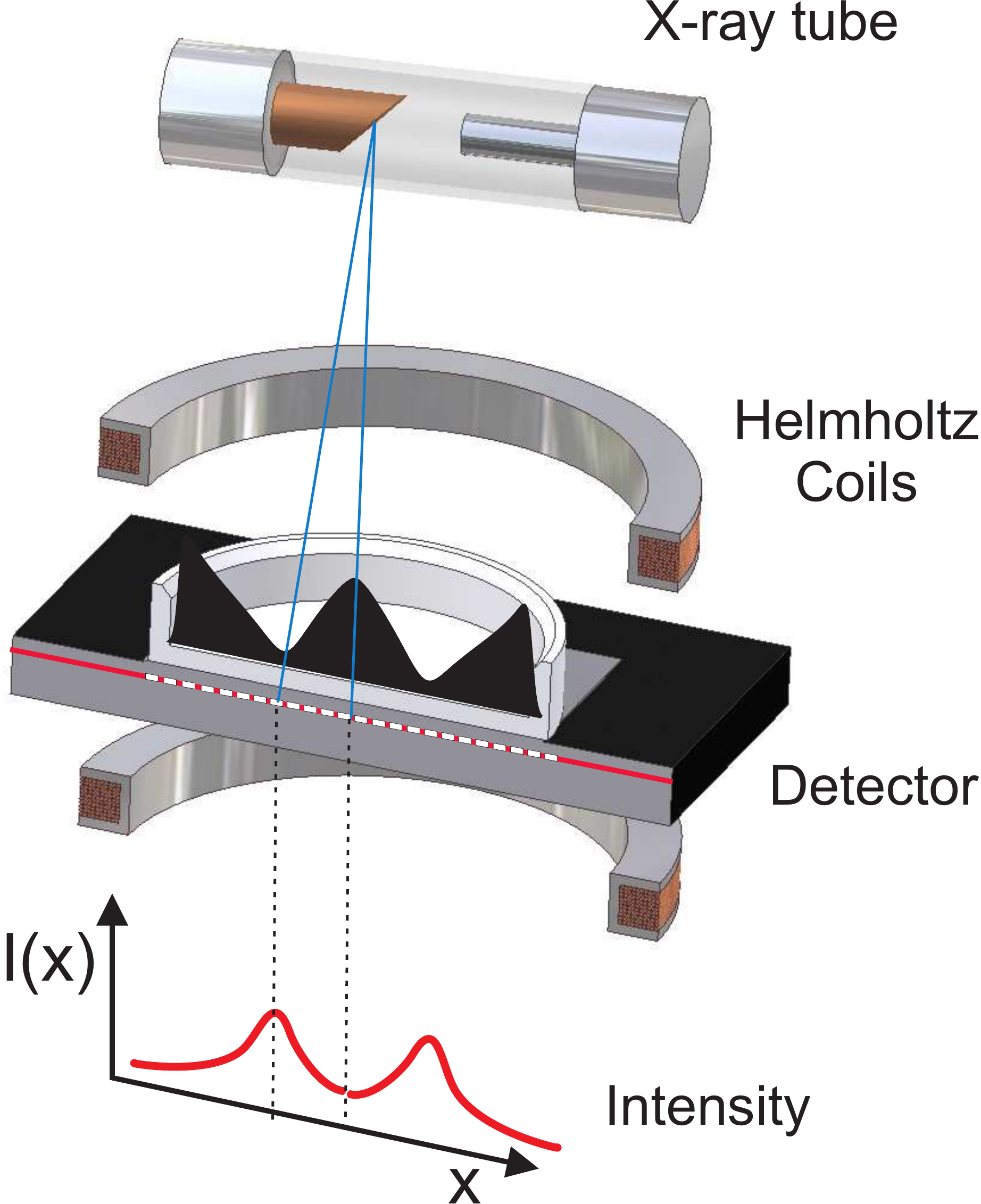} 
  ~\hskip 0.5cm
  \includegraphics[width=5.0cm]{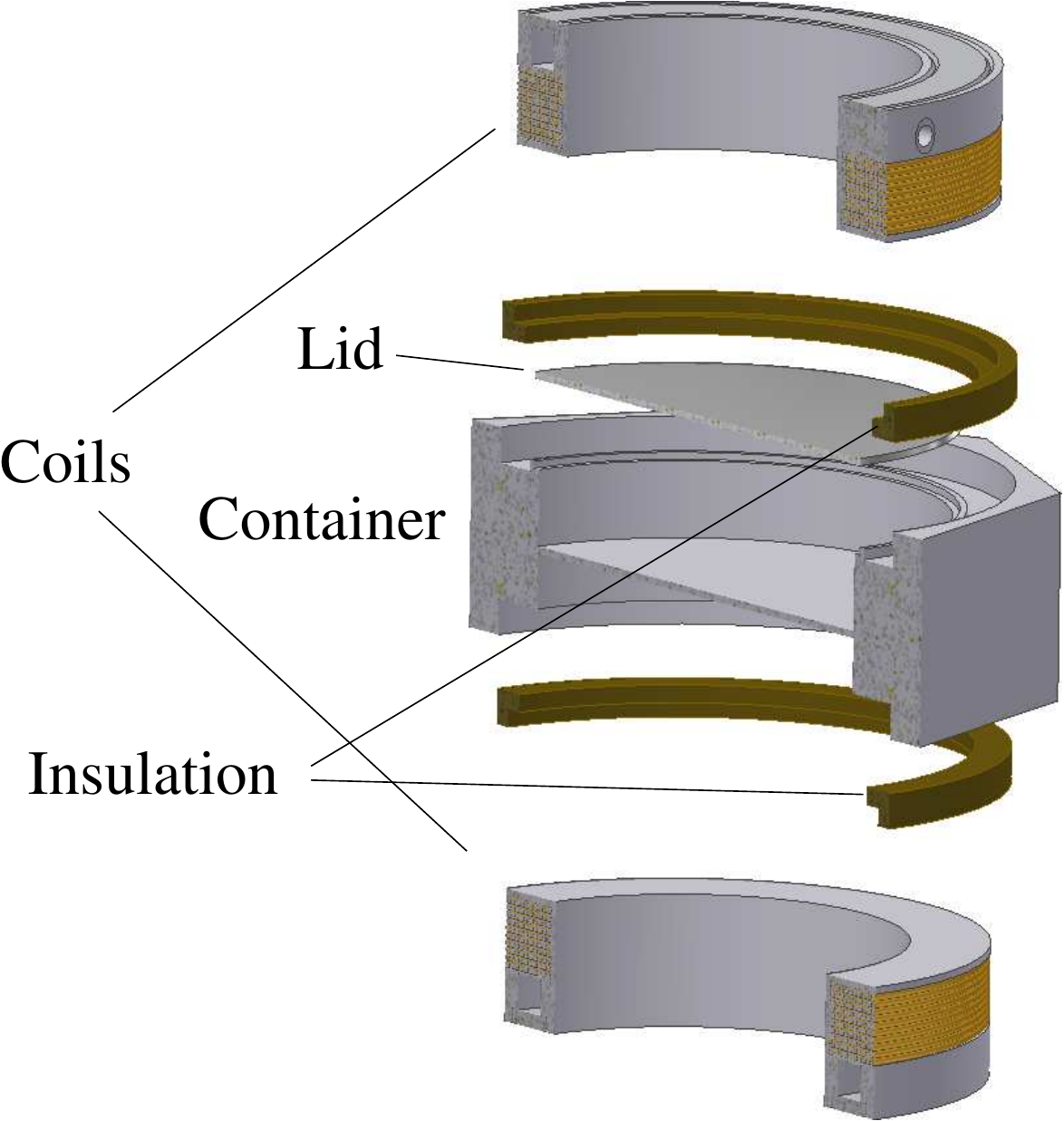}\\[\baselineskip] 
  ~\hskip 0.5 cm (a)~\hskip 5.5 cm (b)\\[\baselineskip]
  \includegraphics[width=7.0cm]{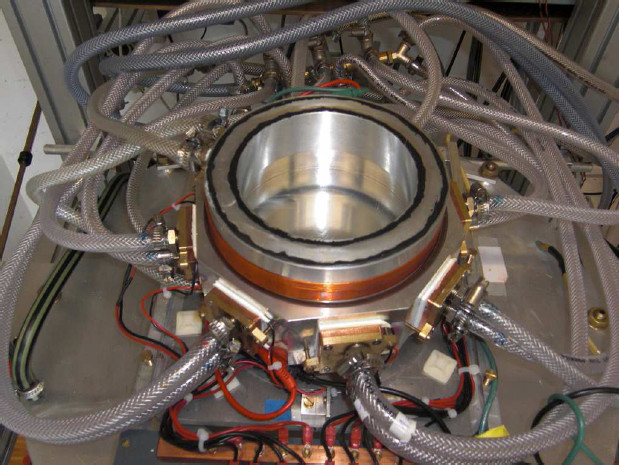}\\[\baselineskip] 
  ~\hskip 0.5cm (c)
  \caption{(Color online) Setup of the apparatus for dynamic measurements of the Rosensweig
  instability. (a) Sketch of the assembled setup of the X-ray machine.
  (b) Detail of the fragmented container with the coils generating the magnetic field.
  (c) Photo of the open container, the upper coil and the water cooled peltier elements.}
  \label{fig:setup}
\end{center}
\end{figure*}

The experimental setup for the measurements of the surface topography
consists of an tailor made X-ray apparatus described in detail before \cite{richter01,gollwitzer07}.
An X-ray point source emits radiation vertically from above through the container filled with the MF. Underneath the container, an X-ray camera records the radiation passing through the layer of MF. The intensity at each pixel of the detector is directly related to the height of
the fluid above that pixel, as sketched in Fig.\ \ref{fig:setup}(a). Therefore, the full surface topography can be reconstructed after calibration \citep{richter01,gollwitzer07}.

The container, which holds the MF sample, is depicted in
Fig.\ \ref{fig:setup}(b). It is a regular octagon machined from
aluminium with a side length of $77\,\mathrm{mm}$ and two concentric inner
bores with a diameter of $140\,\mathrm{mm}$. These circular holes are carved from
above and below, leaving only a thin base in the middle of the vessel with a
thickness of $2\,\mathrm{mm}$. On top of the octagon, a circular aluminium lid is placed,
which closes the hole from above, as shown in Fig.\ \ref{fig:setup}(b).
Each side of the octagon is equipped with a thermoelectric element \texttt{QC-127-1.4-8.5MS} from Quick-Ohm, as shown in Fig.\ \ref{fig:setup}(c). The latter are powered by a $1.2\,\mathrm{kW}$ Kepco \texttt{KLP-20-120} power supply. The hot side of the Peltier elements is connected to water cooled heat exchangers. The temperature is measured at the bottom of the aluminium container with a Pt100 resistor. The temperature difference between the center and the edge of the bottom plate does not exceed $0.1\,$K at the temperature $\theta = 10.0\,\,^\circ$C measured at the edge of the vessel.
A closed loop control, realized using a computer and programmable interface devices, holds $\theta$ constant within $10\,\mathrm{mK}$.

The container is surrounded by a Helmholtz-pair-of-coils, thermally isolated
from the vessel with a ring made from the flame resistant material \texttt{FR-2}.
The size of the coils is adapted to the size of the vessel in order to introduce a "magnetic ramp" at the edge of the vessel. This technique, as described more detailed in Ref.\,\cite{knieling10}, serves to minimize distortions by compensating partly the jump of the magnetization at the container edge.
Filling the container to a height of $5\,\mathrm{mm}$ with ferrofluid enhances the magnetic induction in comparison with the empty coils
for the same current $I$. Therefore $B(I)$ is measured immediately beneath the
bottom of the container, at the central position, and serves as the control
parameter in the following.

\subsection{\label{subsec:fluid} Characterization of the ferrofluid}
The vessel is filled with the commercial magnetic fluid APG\,E32 from Ferrotec Co. up to a hight of 5\,mm.
The material parameters of this MF are listed in Tab.\ \ref{tab:apge32_parameters}. The density was measured using a \texttt{DMA~$4100$} density meter from Anton Paar.
The surface tension was measured using a commercial ring tensiometer (Lauda \texttt{TE~$1$}) and a pendant drop method (Dataphysics \texttt{OCA~$20$}). Both methods result in a surface tension of $\sigma=(31\pm0.5)\,\mathrm{mN\,m^{-1}}$, but when the liquid is allowed to rest for one day,
$\sigma$ drops down to $(25\pm0.5)\,\mathrm{mN\,m^{-1}}$. This effect, which is not
observed in similar, but less viscous magnetic liquids like the one used in Ref.\
\cite{gollwitzer09}, gives a hint that the surfactants change the surface tension at least on a longer time scale, when the surface is changed.
Since indeed the pattern formation experiments do change the surface during the measurements,
the uncertainty of the surface tension is $\approx5\,\mathrm{mN\,m^{-1}}$, as given in Tab.\ \ref{tab:apge32_parameters}.
\begin{table*}[htbp]
\caption{Material properties of the magnetic fluid  APG\,E32 (Lot~G090707A) from Ferrotec Co.}
\begin{ruledtabular}
\begin{tabular}{lcdcc}
Quantity                                        &                & \textnormal{Value}    & Error       & Unit                 \\ \hline
Density	at $10\;\,^\circ\mathrm{C}$               & $\rho$         & 1168.0 & $\pm 1$     & $\mathrm{kg\,m^{-3}}$ \\
Surface tension at $10\;\,^\circ\mathrm{C}$          & $\sigma$       & 30.9   & $\pm 5$     & $\mathrm{mN\,m^{-1}}$ \\
Viscosity at $10\;^\circ\mathrm{C}$             & $\eta$         & 4.48   & $\pm 0.1$   & $\mathrm{Pa\,s}$\\
Saturation magnetization                        & $M_\mathrm{S}$ & 26.6   & $\pm 0.8$   & $\mathrm{kA\,m^{-1}}$ \\
Initial susceptibility at $10\;^\circ\mathrm{C}$& $\chi_0$       & 3.74   & $\pm 0.005$ & \\
Fit of $M(H)$ with the model by Ref.\cite{ivanov01} &                &        &             & \\
~~~Exponent of the $\Gamma$-distribution        & $\alpha_\mathrm{\Gamma}$ & 3.8    & $\pm 1$     & \\
~~~Typical diameter of the bare particles       & $d_0$          & 1.7    & $\pm 0.2$   & $\mathrm{nm}$ \\
~~~Volume fraction of the magnetic material     & $\phi$         & 5.96   & $\pm 0.2$   & \% \\
Fit of $\eta(H)$ with the model by Ref.\cite{shliomis1972}& & & &\\
~~~Mean diameter of the bare particle           & $d_\mathrm{m}$            & 15     &             &
$\mathrm{nm}$ \\
~~~Volume fraction of the magnetic material     & $\phi$         & 21.4   & $\pm 0.2$   & \% \\
Critical induction for a semi-infinite layer \cite{rosensweig1985} & $B_\mathrm{c,\,theo,\,lin,\,\infty}$         & 10.5  & $\pm 0.1$   & $\mathrm{mT}$\\
\end{tabular}
\end{ruledtabular}
\label{tab:apge32_parameters}
\end{table*}

\begin{figure}[htbp]
\begin{center}
\includegraphics[width=0.8\linewidth]{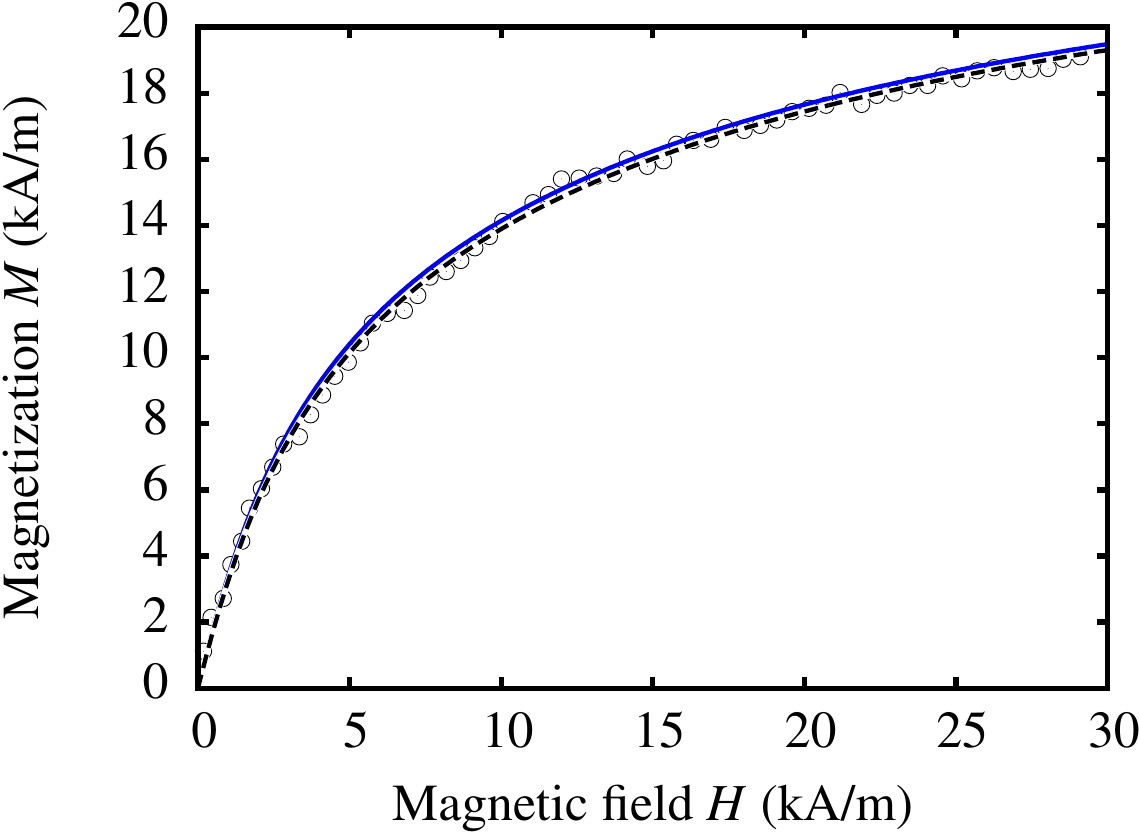}    
 \caption{(Color online) Magnetization curve of the magnetic fluid APG\,E32 measured with a fluxmetric magnetometer. The symbols show the measured data at $\theta=20\,^\circ\mathrm{C}$. The black dashed line is a fit with the model by Ref.\,\protect\cite{ivanov01}. The blue solid line marks an extrapolation to $\theta = 10\,^\circ$C according to this model.
}
\label{fig:magkurve}
\end{center}
\end{figure}

\paragraph{Magnetization curve}
The magnetization has been determined using a fluxmetric magnetometer
(Lakeshore Model~480) constructed to deal with larger samples of high viscosity at a temperature of $\theta=20\,^\circ\mathrm{C}$. Figure\,\ref{fig:magkurve} shows the data, which have been fitted by the modified mean field model of second order \cite{ivanov01}, marked by the dashed black line.
For a comparison with the pattern formation experiments performed at $\theta=10\,^\circ\mathrm{C}$, this curve is extrapolated  utilizing this model (blue line). The deviation between both curves is tiny, which was corroborated with a vibrating sample magnetometer (Lakeshore VSM 7404) at $\theta=20\,^\circ\mathrm{C}$ and $10\,^\circ\mathrm{C}$. Note that the VSM offers the advantage that it can be tempered, but has a lower resolution in comparison to the fluxmetric device because of the smaller sample volume.

To take into account the nonlinear $M(H)$, an effective susceptibility $\bar\chi_\mathrm{H}$ is defined by a geometric mean
\begin{equation}
1+\bar\chi_\mathrm{H}
=\sqrt{(1+\chi_\mathrm{ta})(1+\chi_\mathrm{ch})},
\label{eq:chi.eff}
\end{equation}
with the tangent susceptibility $\chi_\mathrm{ta}=\partial M/\partial H$ and the chord susceptibility $\chi_\mathrm{ch}=M/H$ \cite{zelazo69}. For any field $H$ the effective susceptibility $\bar\chi_\mathrm{H}$ can be evaluated, when the magnetization $M(H)$ curve is known.

\paragraph{Viscosity}
The viscosity $\eta$ deserves special attention for the experiments, as it
influences the time scale of the pattern formation. It has been measured in a
temperature range of $-5\;^\circ\mathrm{C}\leq \theta \leq 20\,^\circ\mathrm{C}$,
using a commercial rheometer (MCR\,301, Anton Paar) with a shear cell featuring
a cone-plate geometry. At room temperature, the magnetic fluid with a viscosity of $\eta=2\,\mathrm{Pa\,s}$ is $2000$ times more viscous than water. The value of $\eta$ can be increased by factor of $9$ when the liquid is cooled to $-5\;^\circ\mathrm{C}$. The temperature dependent viscosity data can be nicely fitted with the well-known Vogel-Fulcher law
\cite{rault00}
\begin{equation}
\eta=\eta_0\exp\left(\frac{\psi}{\theta-\theta_0}\right),
\label{eq:vogelfulcher}
\end{equation}
with $\eta_0=0.48\,\mathrm{mPa\,s}, \psi=1074\,\mathrm{K},$ and
$\theta_0=-107.5\,^\circ\mathrm{C}$, as described in detail in Ref.\,\cite{gollwitzer10}.
For the present measurements, we chose a temperature of
$\theta=10\,^\circ\mathrm{C}$, where the viscosity amounts to
$\eta=4.48\,\mathrm{Pa\,s}$ according to Eq.\ (\ref{eq:vogelfulcher}).

\paragraph{Magnetoviscosity}
\begin{figure}[htbp]
\begin{center}
\includegraphics[width=0.8\linewidth]{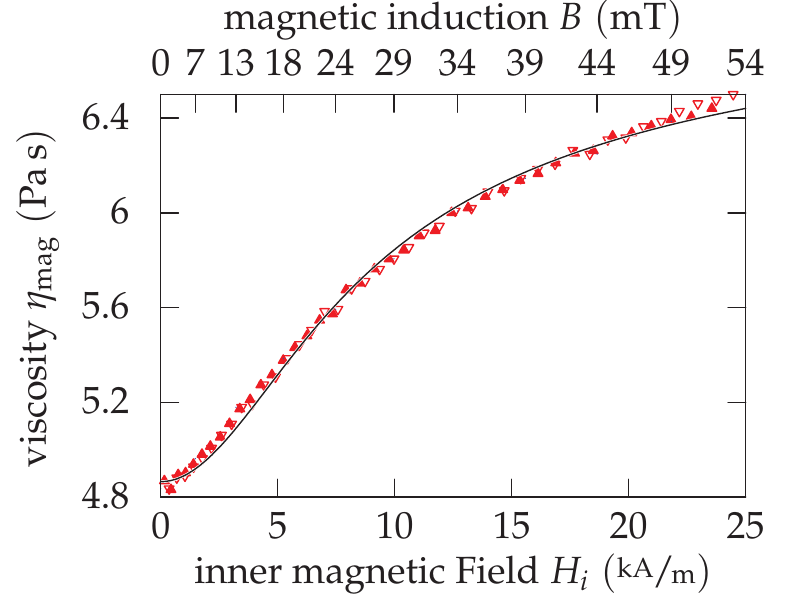}    
\caption{(Color online) The magnetoviscosity of the ferrofluid APG\,E32  versus the inner field $H_\mathrm{i}$ for a shear rate of $\dot{\gamma}=1\;\mathrm{s^{-1}}$. The $\blacktriangle$ ($\bigtriangledown$) mark measurements for increasing (decreasing) $H_\mathrm{i}$ and the solid line is a fit by Eq.\ (\ref{eq:shliomis1972}).
The upper abscissa displays the applied magnetic induction $B$ measured in the air gap beneath the magnetorheological cell.}
\label{fig:mag.viscosity}
\end{center}
\end{figure}

The growth and decay of ferrofluidic spikes takes place in a magnetic field, which is known to alter the viscosity. Furnishing the rheometer with the magnetorheological device MRD 170-1T from Anton Paar we exemplary measure the magnetoviscous behaviour for a shear rate of $\dot{\gamma}=1\, \mathrm{s^{-1}}$. We use a plate-plate configuration with a gap of $300\,\mathrm{\mu m}$, where the upper plate has a diameter of 20\,mm. Figure\,\ref{fig:mag.viscosity} displays the measured data together with a fit by
\begin{equation}
\eta(\alpha) = \eta|_{\alpha=0} + \eta_\mathrm{r}(\alpha) = \eta|_{\alpha=0} + \frac{3}{2} \Phi_\mathrm{h} \eta \frac{\alpha - \tanh \alpha}{\alpha + \tanh \alpha} \langle \sin^2\beta \rangle,
\label{eq:shliomis1972}
\end{equation}
which describes the magnetoviscosity according to Shliomis \cite{shliomis1972}.
Here $\alpha = \mu_0 M_\mathrm{d} V \cdot H_\mathrm{i}/(k_\mathrm{B}T)$, denotes the ratio between the magnetic energy of the dipole in the field $H_\mathrm{i}$ and the thermal energy $k_\mathrm{B}T$, where $M_\mathrm{d}=446\, \mathrm{kA/m}$ is the domain magnetisation of saturated magnetite \cite{rosensweig85_book}, and $V$ the magnetic active volume. Moreover
$\eta|_{\alpha=0}$ captures the viscosity without a magnetic field, $\eta_\mathrm{r}$ the additional rotational viscosity due to the presence of the magnetic field $\vec{H_\mathrm{i}}$ in the ferrofluid, and $\Phi_\mathrm{h}$ is the hydrodynamic volume fraction of the magnetite particles. The brackets $\langle \dots\rangle$ indicate a spatial average over the inclosed quantity. Note that in case of Fig.\,\ref{fig:mag.viscosity} the angle $\beta$ between $\vec{H_\mathrm{i}}$ and the vorticity of the flow is $90^o$. For the fit the internal field was obtained via solving $H_\mathrm{i}=B/\mu_0 - D\cdot M(H_\mathrm{i})$, assuming a demagnetization factor of $D = 1$. The fit yields a hydrodynamic volume fraction of $\Phi_\mathrm{h} =  43.5 \pm 0.1 \%$ and $\alpha / H_\mathrm{i}= 256 \pm 2 \cdot 10^{-6}\;\mathrm{m/A}$. From $V$ one estimates a mean diameter of $d_\mathrm{m}=15\,\mathrm{nm}$ for the magnetic particles. This is almost a factor of ten larger than $d_0 = 1.7\,\mathrm{nm}$ obtained from the magnetisation curve (cf.\,table \ref{tab:apge32_parameters}).
Assuming a spherical layer of oleic acid molecules of thickness $\delta=2\,\mathrm{nm}$ around the magnetic particles \cite{rosensweig85_book}, the volume fraction of the magnetic active material is $\Phi = \Phi_\mathrm{h}\cdot (1 + 2 \delta /d_\mathrm{m})^{-3} = 21.4\%$. This is more than three times larger than the value obtained via the magnetisation curve (cf.\,table \ref{tab:apge32_parameters}).
The elevated values of $d$ and $\Phi$ may be a consequence of magnetic agglomerates, which are not taken into account by Eq.\,(\ref{eq:shliomis1972}).

\begin{figure}[htbp]
\begin{center}
  \includegraphics[width=0.8\linewidth]{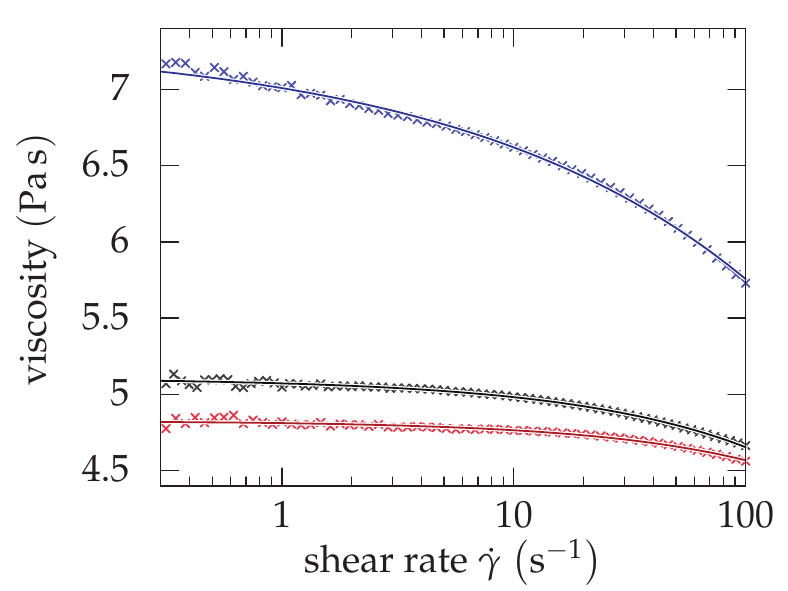}    
  \caption{(Color online) The viscosity of the magnetic fluid APG\,E32
versus the applied shear rate $\dot{\gamma}$ for $B=0\,\mathrm{mT}$ (red), $11.4\,\mathrm{mT}$
(black), and $114\,\mathrm{mT}$ (blue). The crosses mark the measured data (for clarity only every 5th data point is shown), whereas the solid lines display fits by Eq.\,(\ref{eq:sisko}).}
\label{fig:shear.thinning}
\end{center}
\end{figure}

To test the flow behaviour of the ferrofluid, the viscosity was measured versus the shear rate for three exemplary magnetic inductions, as presented in Fig.\,\ref{fig:shear.thinning}.
All curves exhibit a decay of the viscosity for increasing $\dot{\gamma}$, i.e.~shear thinning
which is typical for dispersions \cite{tanner2000}.
For a quantitative description of this effect the measured data are fitted by the Sisko equation \cite{sisko1958}
\begin{equation}
\eta(\dot{\gamma}) = k \dot{\gamma}^{n-1} + \eta_0.
\label{eq:sisko}
\end{equation}
adapted to the limit $\dot{\gamma}\rightarrow 0\,\mathrm{s^{-1}}$, where $\eta \rightarrow \eta_0$. Moreover $k$ denotes a factor and $n$ a scaling exponent. Table \ref{tab:sisko} displays the fitting parameters obtained for the three inductions.
\begin{table}[btp]
\caption{The parameters obtained by fitting Eq.\,(\ref{eq:sisko}) to the experimental data.}
\begin{ruledtabular}
\begin{tabular}{lccc}
$B$ (mT) & $k$ ($\mathrm{Pa\cdot s^{1-1^{n-1}}}$)   &  $n$   & $\eta_0$ (Pa\,s)      \\ \hline
0        & -0.015 $\pm$ 0.001  & 1.62  $\pm$ 0.02   & 4.826  $\pm$ 0.002             \\
11,4     & -0.035 $\pm$ 0.001  & 1.559 $\pm$ 0.006  & 5.107  $\pm$ 0.002             \\
114      & -0.319 $\pm$ 0.001  & 1.346 $\pm$ 0.004  & 7.328  $\pm$ 0.009             \\
\end{tabular}
\end{ruledtabular}
\label{tab:sisko}
\end{table}
Under increase of $B$ most prominently the factor $k$ is varying. For $B=0\,\mathrm{mT}$
$k$ is tiny and we have an almost Newtonian liquid. The factor $k$ doubles at $B=11.4\,\mathrm{mT}$, and eventually enlarges by a factor of ten at at the tenfold value of $B=114\,\mathrm{mT}$. At the same time $\eta_0$ does not even double.
This quantitative description is in agreement with the increasing decay of the curves in Fig.\,\ref{fig:shear.thinning}. The deepening of shear thinning with $B$ has been attributed to the formation of chains and agglomerates of magnetic particles in the field, and their subsequent destruction under shear. Chains have been uncovered by transmission electron microscopy \cite{shliomis74,klokkenburg2006}, and their desctruction has been studied in magnetorheology  \cite{odenbach1998}. For a review see, e.g., Ref.\,\cite{odenbach2009}.

To conclude this paragraph, both, the fit of the magnetoviscous behaviour as well as the shear thinning are indicating that agglomerates of magnetic particles are emerging in the field. In this way the faint non-Newtonian behaviour of the suspension which is already present at zero induction may be enhanced considerably in the field and may cause unexpected dynamics.

\subsection{\label{subsec:measurements}Measurement protocol}

\begin{figure}[tbp]
\includegraphics[width=0.8\linewidth]{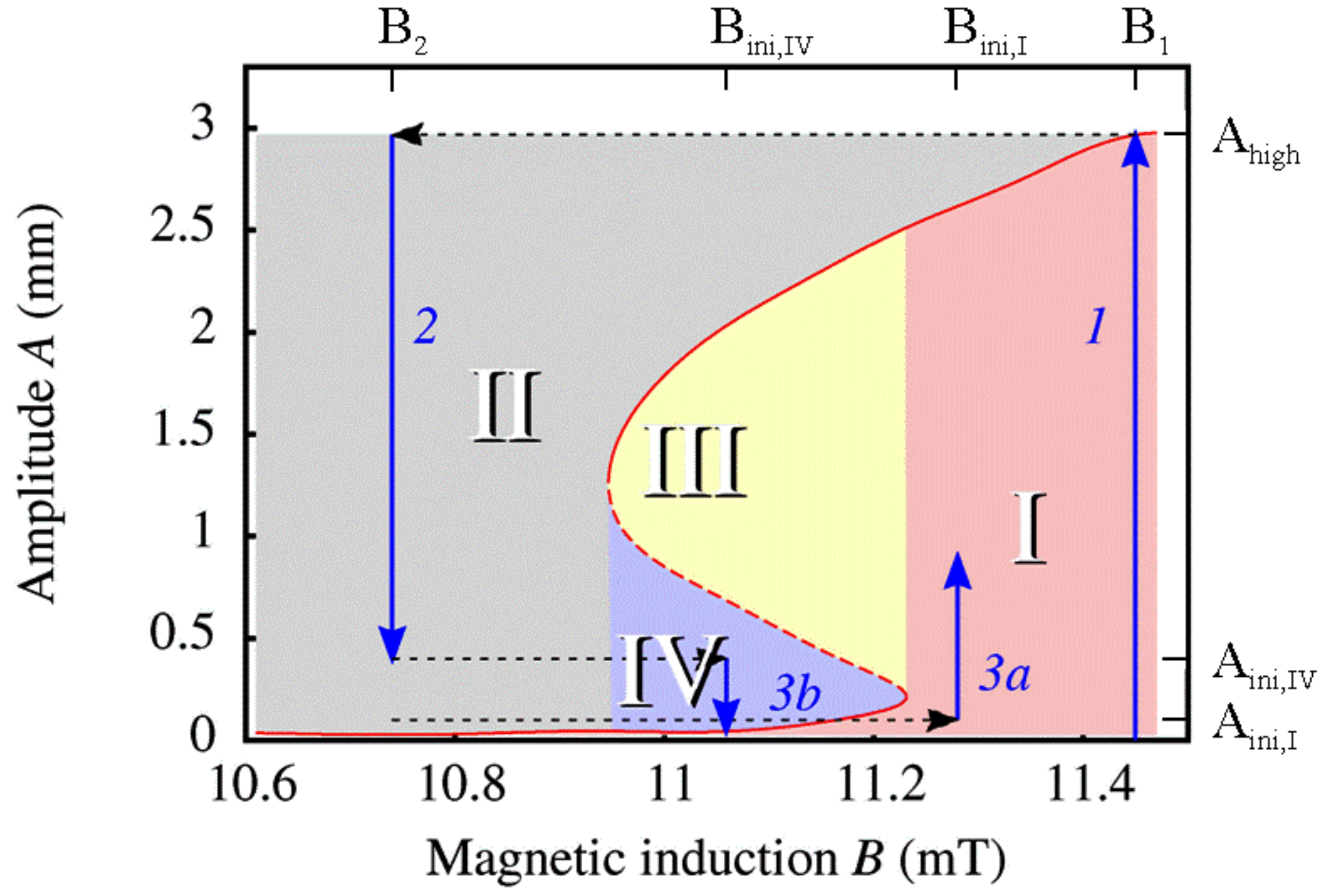}   
\caption{(Color online) The three step measurement protocol for the growth and relaxation measurements.
Dotted arrows indicate jumps of the magnetic inductions. Blue arrows denote the
path of the system during the growth or relaxation phases. Note the different axis labels at the left and at the right,
respectively, as well as at the bottom and at the top, respectively.
A movie of such a process is available at \cite{gollwitzer2007}.}
\label{fig:frozenprotocol}
\end{figure}

Figure~\ref{fig:frozenprotocol} displays the measurement protocol on the basis
of the bifurcation diagram, measured in Ref.\,\cite{gollwitzer10}. The static pattern amplitude of the Rosensweig instability in our fluid is indicated by the red line. When the system is set onto
an arbitrary initial point $(B_\text{ini}, A_\text{ini})$ in this diagram, and the magnetic induction $B$ is kept constant, the amplitude $A$ increases or decreases monotonically, until the
system reaches the stable equilibrium (solid red line). The direction of the
change of $A$ depends on the region, where $(B_\text{ini}, A_\text{ini})$ is situated -- in the
regions I and III in Fig.\,\ref{fig:frozenprotocol}, $A$ increases, and in
regions II and IV, the amplitude decreases with time.

\begin{figure}
\begin{minipage}{0.5\linewidth}
\centering
\includegraphics[width=\linewidth]{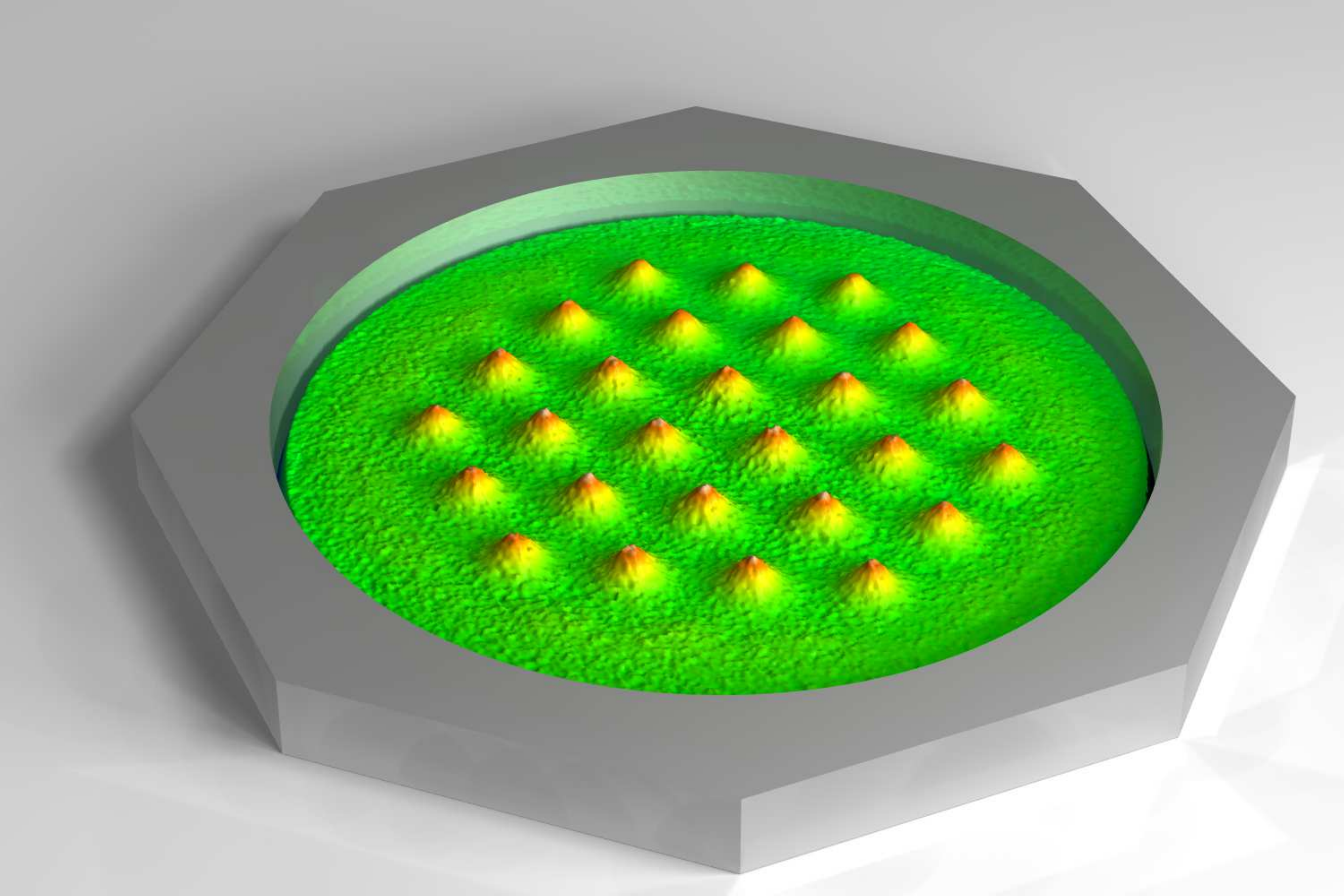} 
(\textit{a})
\end{minipage}\hfill
\begin{minipage}{0.06\linewidth}
\centering
\includegraphics[width=\linewidth]{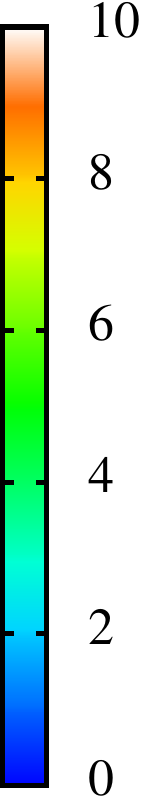} 
(mm)
\end{minipage}\hfill
\begin{minipage}{0.33\linewidth}
\centering
\includegraphics[width=\linewidth]{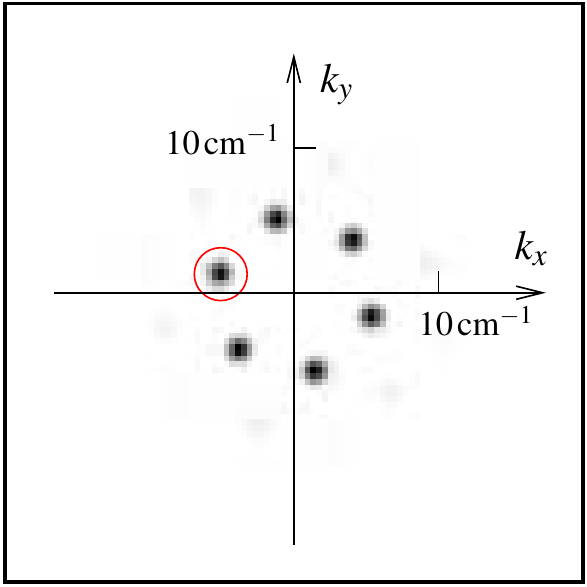} 
(\textit{b})
\end{minipage}
\caption{The final pattern at $B_1=11.45\,\mathrm{mT}$.
Chart (\textit{a}) displays a reconstruction of the surface in real space. The outer
dimension of the container is not to scale. The color code gives the height of the liquid surface
above ground in mm.
The pattern amplitude is determined from the corresponding power spectrum shown in~(\textit{b}) by
the total power in the encircled mode.
The data are taken from Ref.\,\protect\cite{gollwitzer10}.}
\label{fig:wachstum_fourier}
\end{figure}

In order to push the system to an arbitrary initial location $(B_\text{ini},
A_\text{ini})$, a three-step measurement protocol is employed. The first step
(path $\uparrow$\textit{1}) is always a relaxation of the pattern in region~I at the
overcritical induction $B_1=11.45\,\mathrm{mT}$ for
$\tau_1=60\,\mathrm{s}$, to reach the high amplitude of $A_\text{high}=2.98\,\mathrm{mm}$ at that point.
The corresponding pattern is shown in Fig. \ref{fig:wachstum_fourier}.
Then the magnetic induction is quickly reduced to the value $B_2=10.74$mT, and
the resulting dynamics is observed (path $\downarrow$\textit{2}), until the desired starting
amplitude ($A_\text{ini,II}$, $A_\text{ini,IV}$, or $A_\text{ini,I}$) is reached after a period $\tau_2$. To start with this pattern at arbitrary inductions in the regimes II, IV, I
the induction is then quickly raised to the desired value $B_\text{ini}$.
Then we record the pattern evolution along the path $\uparrow$\textit{3a} or $\downarrow$\textit{3b} in region II, IV, and I, respectively.

We use this detour instead of directly switching the
magnetic induction from zero to $B_\text{ini}$ in order to establish the identical pattern
in all regions. Coming from a perfectly flat surface, the pattern would have
additional degrees of freedom, e.g.\ it could amplify any local disturbance,
resulting in a propagating wave front on the liquid surface \cite{knieling10,cao14}.
The emerging hexagonal pattern would comprise point defects or different
orientations of the wave vectors \cite{gollwitzer06,cao14}.
When we take the detour by the paths $\uparrow$\textit{1}
and $\downarrow$\textit{2}, we seed a regular hexagonal pattern at
$(B_1,A_\text{high})$, and the evolving pattern is likely to be
of the same regularity.

\subsection{\label{subsec:extracting}Extraction of the growth rate}
Next we describe the extraction of the growth rate from the recorded sequence of X-ray frames along the path $\uparrow$\textit{3a} or $\downarrow$\textit{3b}. From each X-ray frame the surface topography is reconstructed  following the procedure described in Ref.\,\cite{gollwitzer07}. As an example,  Fig.\,\ref{fig:wachstum_fourier}(a) displays the resulting surface topography at $(B_1,A_\text{high})$. The amplitude of the pattern is determined in Fourier space, as sketched in Fig.\,\ref{fig:wachstum_fourier}(b). We use a circularly symmetric Hamming window with a radius of 46\,mm \cite{gollwitzer10}. The total power in one of the modes, as marked in Fig.\,\ref{fig:wachstum_fourier}(b) by a red circle, is used to compute the amplitude of the pattern \cite{gollwitzer10}.

\begin{figure}[tbp]
\includegraphics[width=0.8\linewidth]{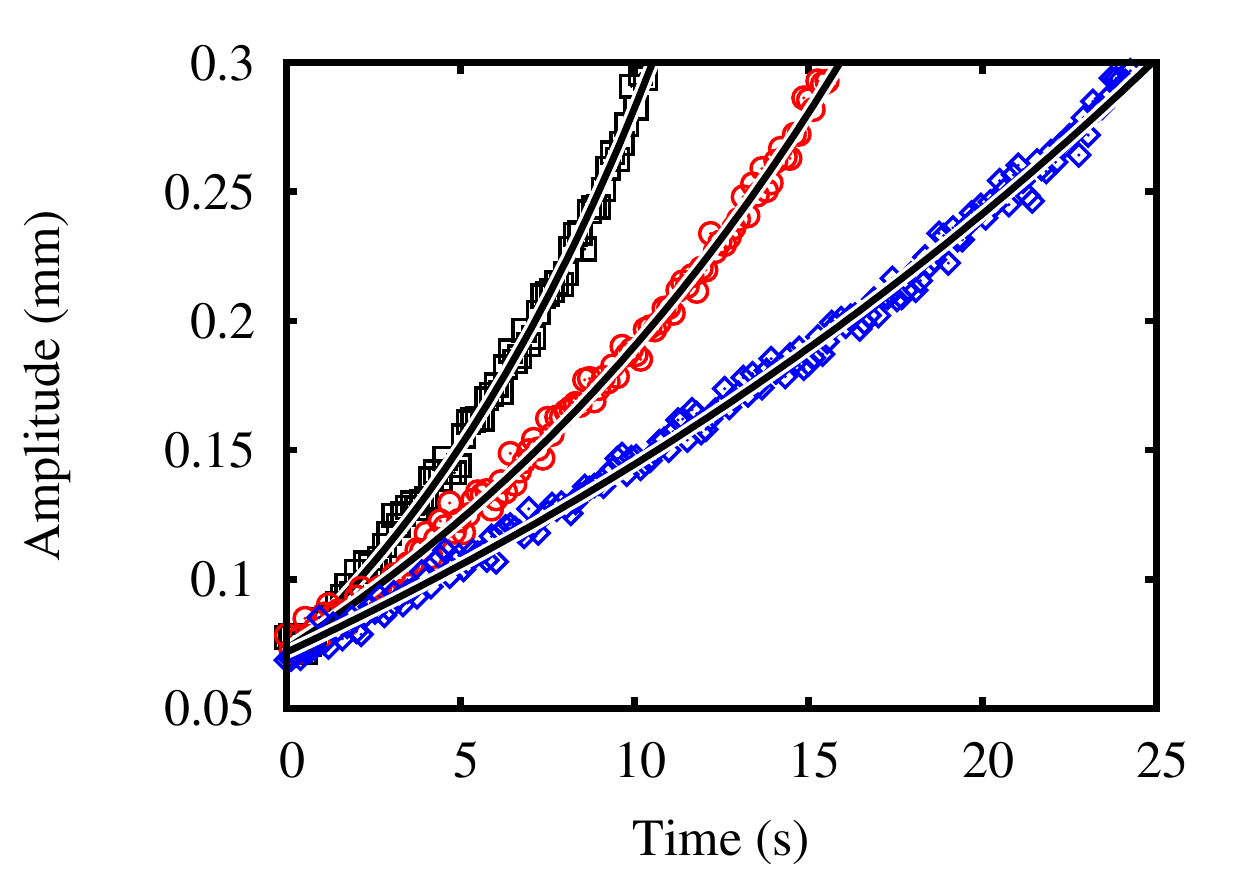}   
\caption{(Color online) Three examples for the growth of the pattern amplitude in region I.
The initial state was prepared with the sequence $B_2 = 10.743\,\mathrm{mT}$, $\tau_2 = 20.000\,\mathrm{s}$.
The three curves have been measured after switching to $B_3=11.455\,\mathrm{mT}$ ({\scriptsize $\square$}), 11.376\,mT ({\scriptsize $\bigodot$}), and 11.323\,mT ({\small $\diamond$}), respectively. The solid black lines denote fits by Eq.\,(\ref{eq:growth.fit}).
}
\label{fig:growth}
\end{figure}

\begin{figure}[tbp]
\includegraphics[width=0.8\linewidth]{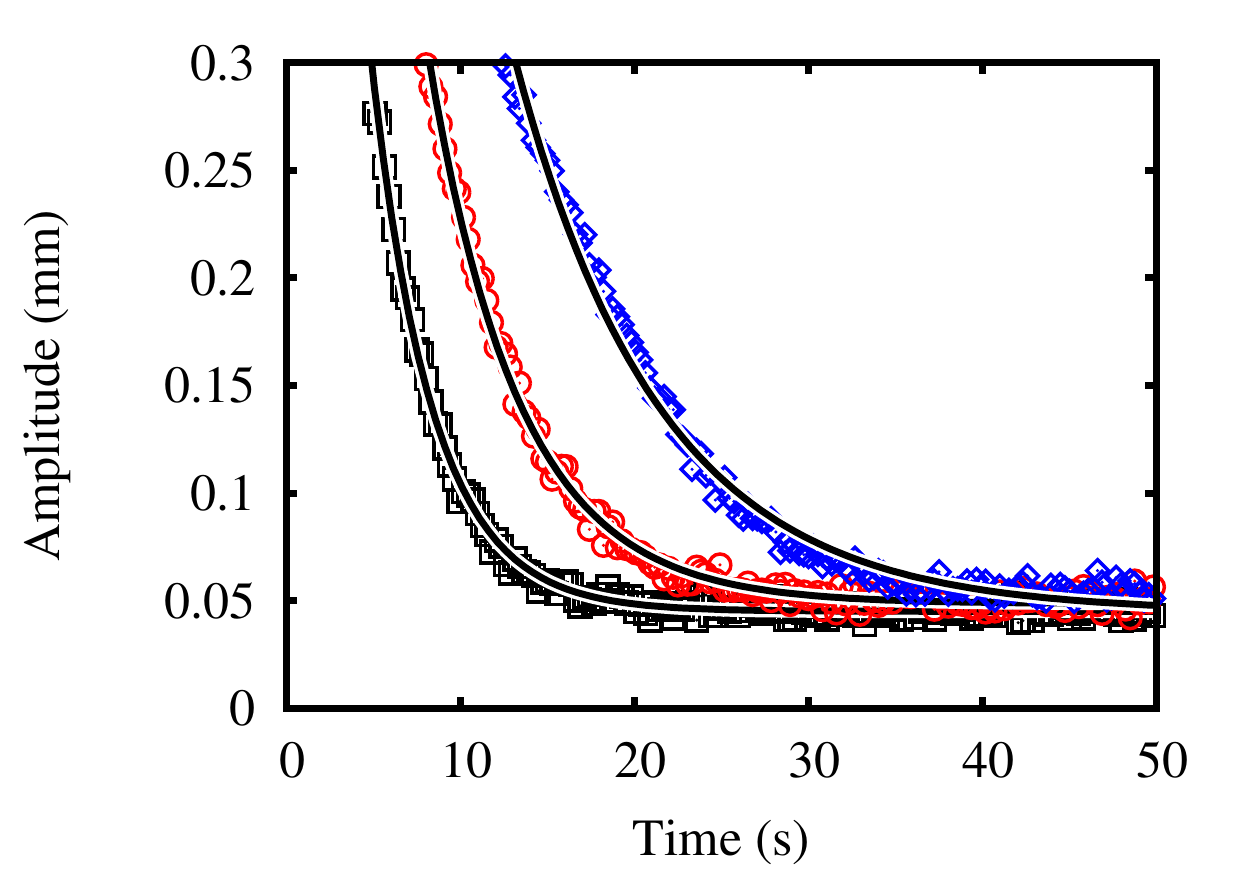}   
\caption{(Color online) Three examples for the decay of the pattern amplitude.
The initial state was prepared with the sequence $B_2 = 10.743\,\mathrm{mT}$, $\tau_2 = 10.000\,\mathrm{s}$.
The three curves have been measured after switching to $B_3=10.888\,\mathrm{mT}$ ({\scriptsize $\square$}), 11.020\,mT ({\scriptsize $\bigodot$}), and 11.059\,mT
({\small $\diamond$}), respectively. The solid black lines denote fits by Eq.\,(\ref{eq:growth.fit}).
}
\label{fig:decay}
\end{figure}

Figure\,\ref{fig:growth} shows three exemplary curves for the growth of the pattern amplitude $A$. With increasing induction $B_3$ (from {\small $\diamond$} via {\scriptsize $\bigodot$} to {\scriptsize $\square$}) the growth increases; likewise Fig.\,\ref{fig:decay} presents three examples for the decay of $A$, where $B_{\rm 3}$ denotes the initial induction $B_{\rm ini}$ after the three steps of the detour procedure.
Remarkably $A$ does not relax to zero, but to a small offset of $A_\mathrm{const}$ which linearly increases from
$32\,\mathrm{\mu m}$ at 10.7\,mT to $34\, \mathrm{\mu m}$ at 10.9\,mT.

A possible explanation are imperfections induced by the lateral container wall, as already observed before, see e.g. Refs.\,\cite{gollwitzer07,friedrich11}. In the present setup special precautions were taken by means of a "magnetic ramp" \cite{knieling10} to minimize such finite size effects. However for an experimental setup with finite aspect ratio they can not be excluded. The fact that $A_\mathrm{const}$ does increase only slightly with $B$ does not contradict this assumption,  because the decay is investigated in the regime II well below $B_\mathrm{c}$. From Fig.\,\ref{fig:frozenprotocol} one clearly unveils that an increase of the imperfection becomes only prominent in the hysteretic regime (IV).

A further explanation would be an inhomogeneous distribution of surfactants at the surface of the MF, which develops after the massive destruction of surface area, which can not be followed up by the diffusion of the surfactants on the surface and into the bulk liquid. A resulting spatial variation of $\sigma$ would lead to surface crests, somehow reminiscent of those observed in Marangoni convection.

A third explanation is an inhomogeneous distribution of magnetic particles due to magnetophoresis \cite{lavrova10} taken place while the field is at $B_1$. We consider this less likely because of the enormous time scales of such a process at the large viscosity of the experimental fluid.

>From each measured curve we extract the linear growth or decay rate by a least square fit with the function
\begin{equation}
A(t) = A_0 \exp{(\omega_2 t)} + A_\mathrm{const},
\label{eq:growth.fit}
\end{equation}
which is taking into account the constant offset. We restrict the fit to the interval $A \in [0.0\,\mathrm{mm}, 0.3\,\mathrm{mm}]$, for which we assume that a linear description is still possible. This is corroborated by exemplary fits in Figs.\,\ref{fig:growth} and \ref{fig:decay} which are marked by solid black lines.

\begin{figure}[tbp]
\begin{center}
\includegraphics[width=0.8\linewidth]{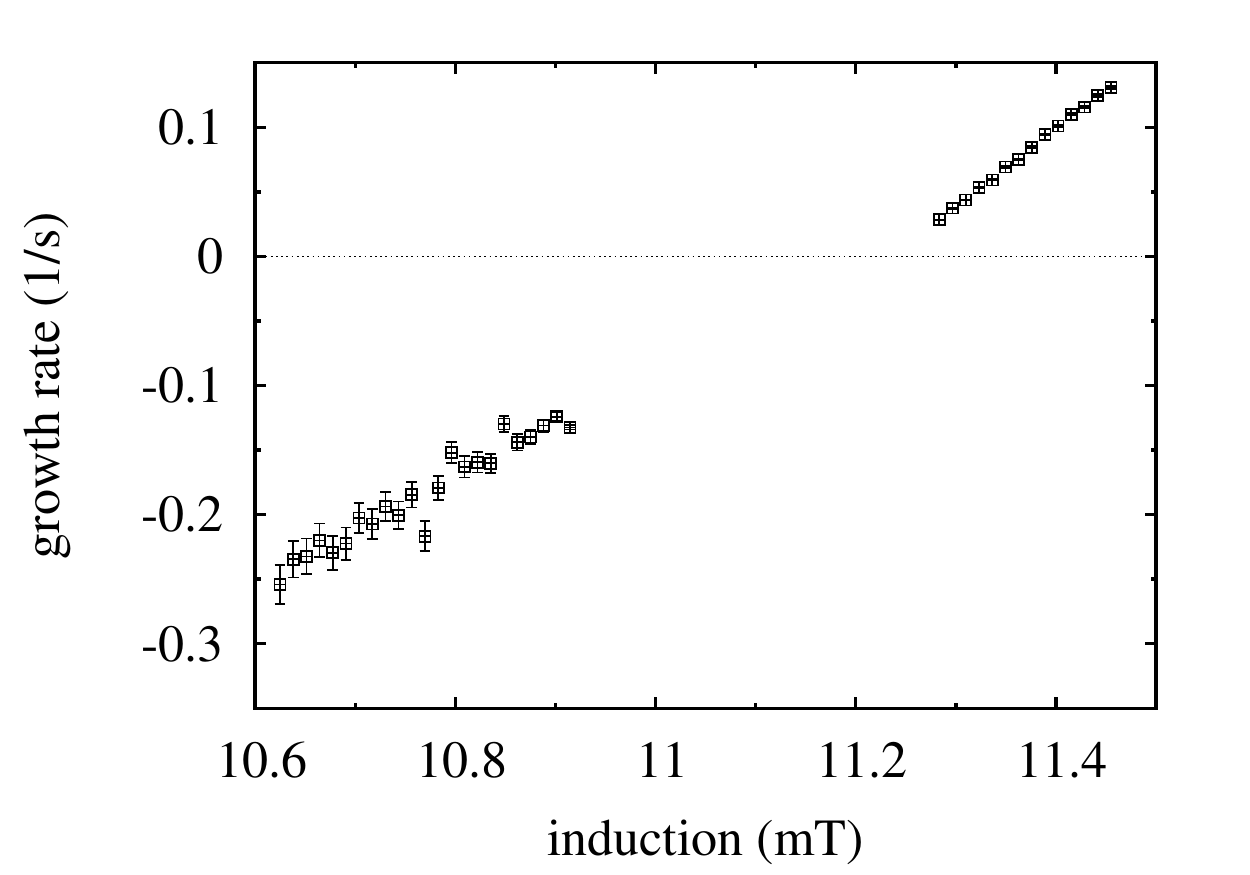}  
  \caption{(Color online) The measured linear growth and decay rates $\omega_2$, respectively,
  of the pattern amplitude as a function of the magnetic induction $B$.}
\label{fig:growthrate_measurement}
\end{center}
\end{figure}

In Fig.\,\ref{fig:growthrate_measurement} we present the extracted growth and decay rate $\omega_2$ versus the applied induction $B=B_3$. Because of the large statistical errors of $\omega_2$ we have refrained to plot the decay rate in the hysteretic regime. The measured values show a monotonous relation with the applied induction, and indicate that a critical value for the magnetic induction of about 11.2~mT exists. Using the material parameters from table\,\ref{tab:apge32_parameters} and an infinite layer thickness yields $B_\mathrm{c, theo, lin,\infty} = 10.5$\,mT \cite{cowley67}.

\section{\label{sec:theory}Theory}
The experimental system, described in Sect.\ \ref{sec:experiment}, is
modeled as a horizontally unbounded layer of an incompressible, nonconducting,
and viscous magnetic fluid subjected to a magnetic field which is perpendicular
to the plain, horizontal, and undisturbed surface. The fluid is bounded from below
by the bottom of a container made of a magnetically insusceptible material and has a
free surface with air above.

According to the linear stability analysis \cite{reimann03}, the pattern amplitude
$A$ can be described by an exponential growth,
$A\sim \mathrm{exp}\left( -\mathrm{i}\tilde\omega t\right)$,
with an exponent $\tilde\omega = \omega_1 + \mathrm{i}\omega_2$, when $A$ is small.
The real part of $-\mathrm{i}\tilde \omega$, $\omega_2$, is called the growth rate and
defines whether the disturbances will grow ($\omega_2>0$) or decay
($\omega_2<0$). The absolute value of the imaginary part of
$-\mathrm{i}\tilde \omega$, $|\omega_1|$, gives the angular frequency of the
oscillations if it is different form zero \cite{reimann03}.

The exponent $\tilde\omega$ follows from the dispersion relation given in Ref.\,\cite{knieling07}
for a layer of MF with the finite depth $h$, a nonlinear magnetization curve $M(H)$,
the surface tension $\sigma$, the density $\rho$, and the kinematic viscosity $\nu$
\begin{widetext}
\begin{eqnarray}
  0 =&
  \displaystyle{
  \frac{\nu^2}{\tilde k\coth(\tilde k h)- k\coth (kh)}
               }
  \biggr(
        \tilde k\left[ 4k^4 +(k^2+\tilde k^2)^2\right]
        \coth(\tilde k h)-k\bigr[ 4k^2\tilde k^2
        +(k^2 +\tilde k^2)^2\bigr]\nonumber \\
      &\times \tanh (kh)-
        \displaystyle{
        \frac{4k^2\tilde k (k^2+\tilde k^2)}
        {\cosh (kh) \sinh (\tilde k h)}
                     }
  \biggr)
        +\tanh (kh) \biggr( g k +\frac{\sigma}{\rho}\,k^3
        -\frac{\mu_0 (1+\bar\chi_\mathrm{H}) M^2}{\rho}
        \Lambda (kh)\,k^2\biggr) \; ,\nonumber \\
  &
\label{eq:disprel}
\end{eqnarray}
where
\begin{equation}
  \Lambda (kh)=
  \displaystyle{
  \frac{ {\rm e}^{kh(1+\bar\chi_\mathrm{H})/(1+\chi_\mathrm{ta})}(2+\bar\chi_\mathrm{H}) -\bar\chi_\mathrm{H}
         {\rm e}^{-kh(1+\bar\chi_\mathrm{H})/(1+\chi_\mathrm{ta})} }
  { {\rm e}^{kh(1+\bar\chi_\mathrm{H})/(1+\chi_\mathrm{ta})}(2+\bar\chi_\mathrm{H})^2 -\bar\chi_\mathrm{H}^2
    {\rm e}^{-kh(1+\bar\chi_\mathrm{H})/(1+\chi_\mathrm{ta})} }
               }
\label{eq:Lambda}
\end{equation}
and
\begin{equation}
  \tilde k = \sqrt{k^2 - \frac{\mathrm{i}\tilde \omega}{\nu} } \; .
\label{eq:ktilde}
\end{equation}
\end{widetext}

The solutions for the dispersion relation in case of a linear magnetization curve were
revised in Ref.\,\cite{lange_mhd03}. The solution space is rather complex, but
the following conclusions can be drawn:
for $k=k_\mathrm{c}$, $\tilde\omega$ is purely imaginary and
the pattern grows or decays exponentially.

\subsection{\label{subsec:scaling_laws}Scaling laws for a nonlinear magnetization}
In the following  we study the generic dependence of the maximal growth rate $\omega_{\rm 2,\,m}$ and the corresponding wave number $k_{\rm m}$ on the nonlinear magnetization of the fluid and its viscosity.
The reason is that $\omega_{\rm 2,\, m}$ and $k_{\rm m}$ characterize the linearly most
unstable pattern. The dispersion relation~(\ref{eq:disprel}) for $\tilde\omega=\mathrm{i}\omega_2$, and an infinitely thick layer, $h\rightarrow\infty$, \cite{salin93}
\begin{align}\label{eq:h_inf_disprel}
\sqrt{1+\frac{\omega_2}{\nu k^2}} &= \left(1+\frac{\omega_2}{2\nu k^2}\right)^2 \nonumber \\
&+\frac{1}{4\rho\nu^2 k^4}\left[
\rho g k +\sigma k^3 - \frac{\mu_0\left(1+\bar\chi_\mathrm{H}\right)}{\left(2+\bar\chi_\mathrm{H}\right)}M^2 k^2
\right]
\end{align}
is written in dimensionless form (indicated by a bar)
\begin{equation}\label{eq:dimless_disprel}
\sqrt{1+\frac{\bar\omega_2}{\bar\nu \bar k^2}} =  \left( 1 + \frac{\bar\omega_2}{2\bar\nu \bar k^2}\right)^2
   + \frac{\bar k + \bar k^3-2 \bar k^2 \bar M^2 \chi_\mathrm{rel}}{4\bar\nu^2 \bar k^4} \; .
\end{equation}
For this result, any length, the time, the kinematic viscosity, and the
magnetization were rescaled to dimensionless quantities using
\begin{subequations}
\label{eq:all_scalings}
\begin{equation}
\bar l = k_\mathrm{c, \infty}\, l = \sqrt{\frac{\rho g}{\sigma}}\,l\; ,
\label{eq:scaling_lenth}
\end{equation}
\begin{equation}
\bar t = \frac{t}{t_\mathrm{c}}=\frac{g^{3/4}\rho^{1/4}}{\sigma^{1/4}}\,t\; ,
\label{eq:scaling_time}
\end{equation}
\begin{equation}
\bar \nu = \frac{g^{1/4} \rho^{3/4}}{\sigma^{3/4}}\, \nu\; ,
\label{eq:scaling_visc}
\end{equation}
and
\begin{equation}
\bar M = \frac{M}{M_\mathrm{c,\infty}} \; ,
\label{eq:scaling_magn}
\end{equation}
where
\begin{equation}
M_\mathrm{c,\infty} = \sqrt{\displaystyle{\frac{2}{\mu_0}}\left( \frac{2+\bar\chi_\mathrm{H_\mathrm{c}}}{1+\bar\chi_\mathrm{H_\mathrm{c}}}\right)\sqrt{ \rho g \sigma} } \;
\label{eq:M.c.infty}
\end{equation}
\end{subequations}
gives the critical magnetisation for a semi-infinite layer of MF.

For finding a scaling law for the growth rate, we differentiate implicitly the dimensionless dispersion relation~(\ref{eq:dimless_disprel}) with respect to $\bar M$, i.e. we determine the slope of the growth rate called $\Gamma$. By taking the limit $\bar M=1$
we find $\Gamma$ in the vicinity of the point of bifurcation,
\begin{equation}\label{eq:Gamma}
    \Gamma=\frac{\partial \bar\omega_2}{\partial \bar M}\biggr|_{\bar M = 1}=\frac{2\chi_\mathrm{rel}}{\bar\nu}\; .
\end{equation}
Inspecting Eq.\ (\ref{eq:Gamma}) one sees that $\Gamma$ is independent of the wave number $\bar k$ and it is a finite constant. Exploiting the latter and using that at the point of bifurcation $\bar\omega_2(\bar M=1)=0$ holds, the following scaling law can be formulated:
\begin{equation}
\label{eq:scaling_all_omega2}
\bar\omega_2=\frac{2\chi_\mathrm{rel}}{\bar\nu}\hat M\; .
\end{equation}

That linear dependence of $\bar\omega_2$ on $\hat M$ is universal and is depicted already in the measured growth rates presented in
Fig.\ \ref{fig:growthrate_measurement}. Since $\Gamma$ scales with $1/\bar\nu$, the slope of the growth rates goes to infinity in the limit of inviscid fluids.
Moreover, for normal magnetic fluids with their rather low viscosity the range of validity of Eq.\ (\ref{eq:scaling_all_omega2}) is bounded by $\bar\nu^2/6$ which
is very small, see third row, third line in Tab.\ \ref{tab:ingredients_growthrate_surfacetension}. Therefore this scaling law is only of limited practical value.

For low viscous fluids it holds that $\omega_2/\left( \nu k^2\right)=\bar\omega_2/\left( \bar\nu \bar{k}^2\right)\gg 1$, see third row, last line in Tab.\ \ref{tab:ingredients_growthrate_surfacetension}.
With the latter inequality, Eq.\ (\ref{eq:dimless_disprel}) simplifies to
\begin{equation}\label{eq:dimless_disprel_low_vis_simple}
    \bar\omega_2^2=-\bar k -\bar k^3 +2\bar k^2 \bar M^2 \chi_\mathrm{rel}
\end{equation}
and one can now determine the slope of the square of the growth rate
\begin{equation}\label{eq:Gamma_low_vis}
    \Gamma_\mathrm{low\,vis}=\frac{\partial\, \bar\omega_2^2}{\partial \bar M}=4\bar k^2 \chi_\mathrm{rel} \bar M\; .
\end{equation}
That scaling law that states that $\bar\omega_2$ scales with the square root of $\bar M$ is of great practical use, as it is shown below.

\begin{table*}[htbp]
\caption{Essential features of a high viscous magnetic fluid like APG\,E32
and those of a low viscous fluid like EMG\,909 associated with
Eq.\ (\ref{eq:Gamma_low_vis}).}
\begin{ruledtabular}
\begin{tabular}{lll}
Quantity                                & high viscous MF                   & low viscous MF \\ \hline
$\nu$ ($\mathrm{m}^2/\mathrm{s}$)       & $\simeq 3.8\times 10^{-3}$        & $\simeq 4.2\times 10^{-6}$ \cite{knieling07} \\
$\bar\nu$                               & $\simeq 32.4.0$                     & $\simeq 2.2\times 10^{-2}$ \cite{knieling07} \\
$\bar\nu^2/6$                           & $\simeq 87.9$                     & $\simeq 7.8\times 10^{-5}$         \\
wave number                             & $k\simeq k_\mathrm{c,\,\infty}$   & $k\simeq k_\mathrm{c,\,\infty}
\left( 1 + \tilde c_3 \hat M + \tilde c_4 \sqrt{\hat M}\;\right)$ \cite{lange01_growth,comment1_PRE15} \\
                                        & $k\simeq 705.7$ $\mathrm{m}^{-1}$ & $640.7\;\mathrm{m}^{-1}\lesssim k \lesssim 1210.9\;\mathrm{m}^{-1}$
\cite{knieling07} \\
$\omega_\mathrm{2}$ ($\mathrm{s}^{-1}$) & $\sim 0.1$                        & $\sim 40$          \\
$\displaystyle{
\frac{\omega_\mathrm{2}}{\nu k^2}=\frac{\bar\omega_\mathrm{2}}{\bar\nu \bar{k}^2}
}$                                      & $\ll 1$                           & $\gg 1$            \\
\end{tabular}
\end{ruledtabular}
\label{tab:ingredients_growthrate_surfacetension}
\end{table*}

In Eq.\ (\ref{eq:dimless_disprel}) two scaled material parameters appear, where
$\chi_\mathrm{rel}$ is a function of the magnetic field,
\begin{equation}\label{eq:chi_rel}
\chi_\mathrm{rel} = \frac{(1+\bar\chi_\mathrm{H})}{(2+\bar\chi_\mathrm{H})}\cdot
                    \frac{(2+\bar\chi_\mathrm{H_\mathrm{c}})}{(1+\bar\chi_\mathrm{H_\mathrm{c}})}
\; ,
\end{equation}
and relates the susceptibility at the field strength $H$ to the one at the critical field
$H_\mathrm{c}$ for the Rosensweig instability.
A step towards the scaling laws is the expansion of $\chi_\mathrm{rel}$
in powers of the scaled distance of the magnetization to the critical value, $\hat M =(M-M_\mathrm{c,\,\infty})/M_\mathrm{c,\,\infty}$,
\begin{equation}
\label{eq:expansion_chi_rel}
   \chi_\mathrm{rel} = 1 + a_\chi\hat M + b_\chi\hat M^2 \; .
\end{equation}
In the following we utilize this simplified description of $\chi_\mathrm{rel}$.

By expanding $\bar M$, $\bar\omega_2$, and $\bar k$ with respect to $\hat M$ around their critical values at the onset of the instability, too,
\begin{subequations}
\label{eq:expansion_M_omega_k}
\begin{equation}
\label{eq:expansion_M}
\bar M  =1 + \hat M\; ,
\end{equation}
\begin{equation}
\label{eq:expansion_omega}
\bar\omega_2 = 0 +\hat\omega_{\rm 2,m} = \alpha\hat M + \tilde\beta \hat M^2 +\tilde\gamma \hat M^3 + \Theta \hat M^4 +\iota \hat M^5 + \mathrm{h.\,o.\,t.}\; ,
\end{equation}
\begin{equation}
\label{eq:expansion_k}
\bar k = 1 + \hat k_{\rm m} = 1+ \tilde\lambda \hat M + \delta \hat M^2 + \epsilon \hat M^3
+\xi \hat M^4 + o \hat M^5 +\mathrm{h.\,o.\,t.}\; ,
\end{equation}
\end{subequations}
and following the procedure outlined in Ref.\,\cite{lange01_growth}, one yields two scaling laws valid up to a scaled magnetization of $\hat M \leq \bar\nu^2/6$,
\begin{widetext}
\begin{align}
\nonumber
\hat\omega_\mathrm{2,\,m} = &\left(\frac{2+a_\chi}{\bar\nu}\right) \hat M + \left(\frac{1+2a_\chi+b_\chi}{\bar\nu} - \frac{3(2+a_\chi)^2}{4\bar\nu^3}\right)\hat M^2\\
\label{eq:hat_omega2_hatM}
&+\left( \frac{a_\chi+2b_\chi}{\bar\nu} - \frac{3(2+a_\chi)(1+2a_\chi+b_\chi)}{2\bar\nu^3} + \frac{5(2+a_\chi)^3}{4\bar\nu^5}\right)\hat M^3 +
\Theta \hat M^4 +\iota \hat M^5 \; ,\\
\nonumber
\hat k_\mathrm{m} = &\left( \frac{3(2+a_\chi)^2}{2\bar\nu^2}\right) \hat M^2 +\left( \frac{3(2+a_\chi)(1+2a_\chi+b_\chi)}{\bar\nu^2}- \frac{11(2+a_\chi)^3}{4\bar\nu^4}\right)\hat M^3\\
\label{eq:hat_k_hatM}
&+\xi \hat M^4 + o \hat M^5 \; .
\end{align}
\end{widetext}
Due to their length the coefficients $\Theta$, $\xi$, $\iota$, and $o$ are given in appendix~\ref{sec:appendix_coeff}.
Both scaling laws show the explicit dependence on the parameters viscosity and magnetization which can be any nonlinear function of $H$. Therefore Eqs.\,(\ref{eq:hat_omega2_hatM}) and~(\ref{eq:hat_k_hatM}) represent the generalization
of the results for a linear law of magnetization, i.e. for $a_\chi=b_\chi=0$, given in
Ref.\,\cite{lange01_growth}.

\begin{figure}
\includegraphics[width=0.8\linewidth]{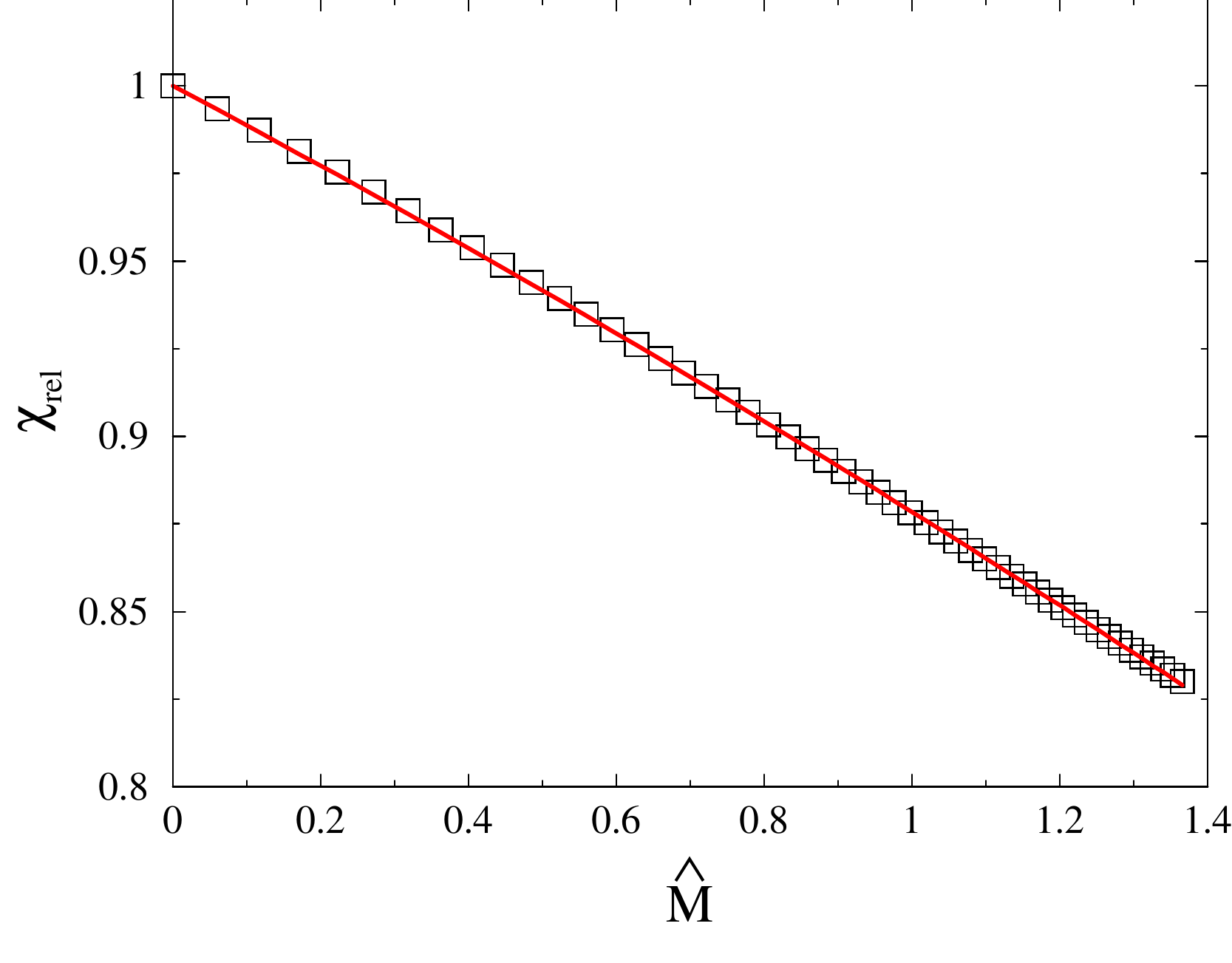}    
  \caption{(Color online) The relative susceptibility $\chi_\mathrm{rel}$ as function of the
  rescaled magnetization $\hat M$ of the measured curve ({\scriptsize $\square$}) and from the second order approximation~(\ref{eq:expansion_chi_rel}) (solid red line).}
  \label{fig:chi_rel}
\end{figure}

To prove the quality of the simplified description of $\chi_\mathrm{rel}$ by Eq.\ (\ref{eq:expansion_chi_rel}),
in Fig.\,\ref{fig:chi_rel} the experimental values of $\chi_\mathrm{rel}$ based on the magnetization curve shown in Fig.\,\ref{fig:magkurve} are determined by fitting that magnetization by the model proposed in Ref.\,\cite{ivanov01}. The solid line represents Eq.\,(\ref{eq:expansion_chi_rel}) with $a_\chi=-0.1118$ and $b_\chi=-0.0097$
resulting in a very good agreement with the experimental data.

\subsection{\label{subsec:lin_growthrate} The finite layer approximation for a highly viscous fluids}
\begin{figure*}[tbp]
\begin{center}
\includegraphics[width=7.9cm]{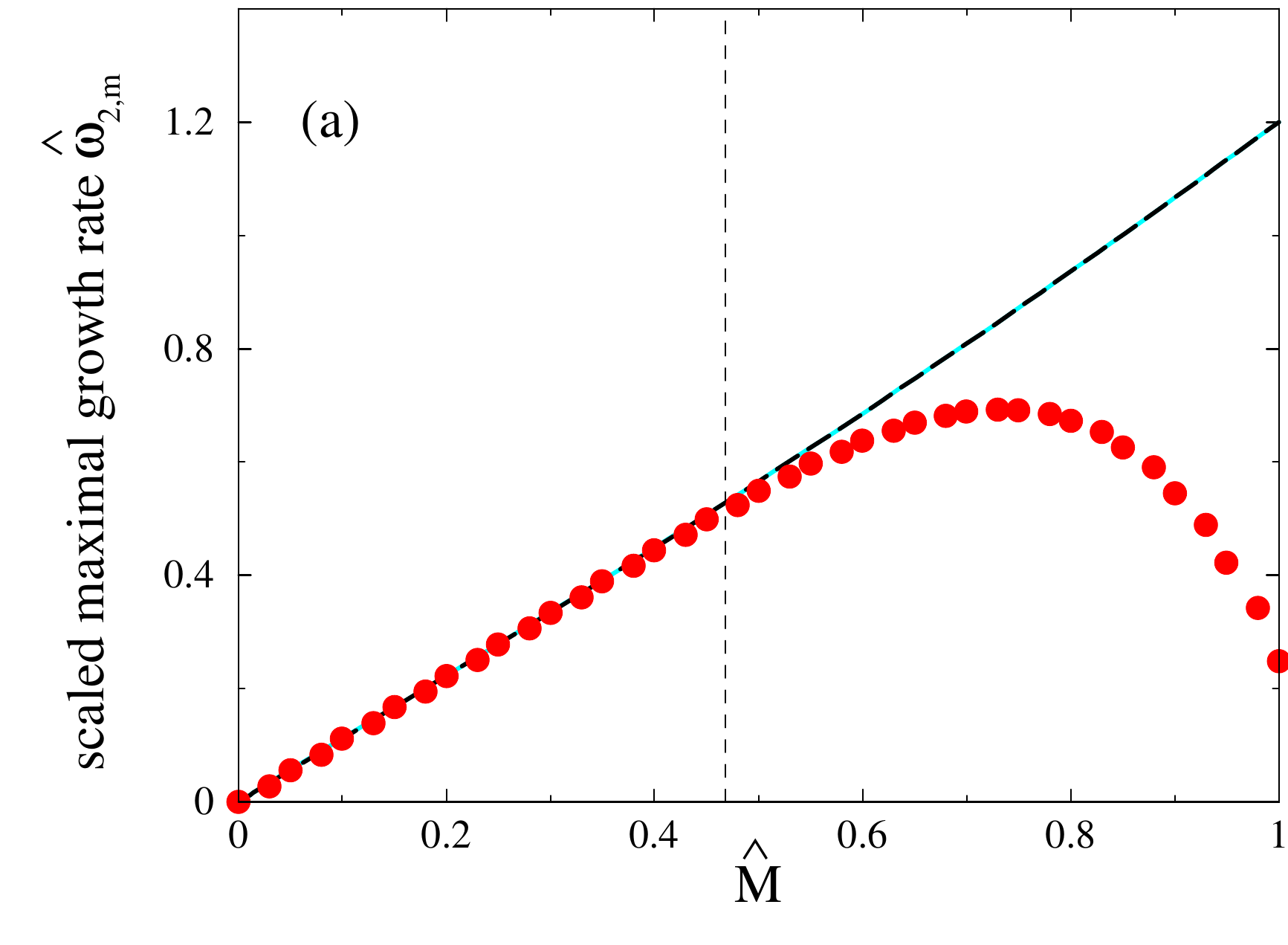} 
~\hskip 0.4 cm
\includegraphics[width=7.9cm]{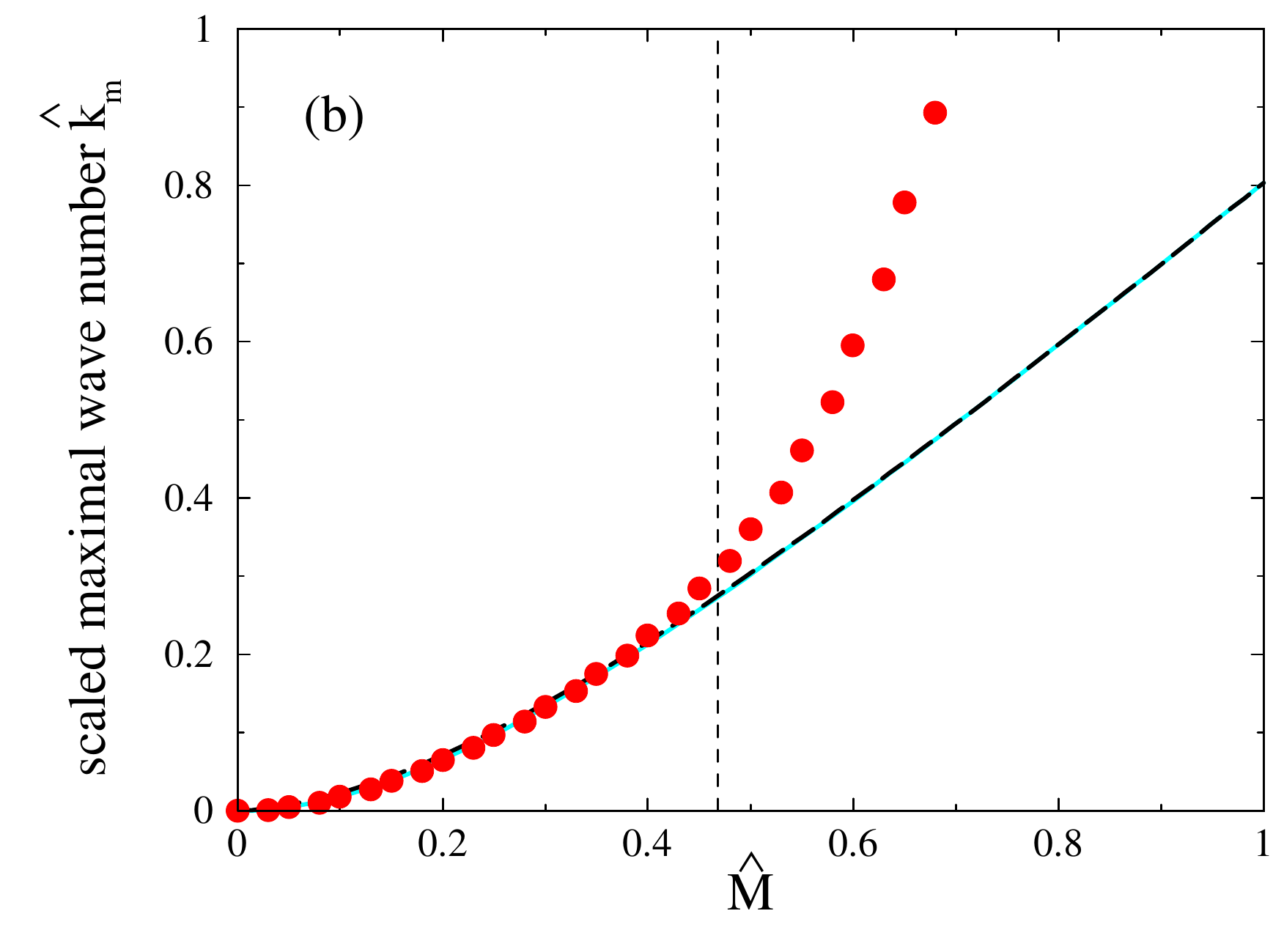} 
\includegraphics[width=7.9cm]{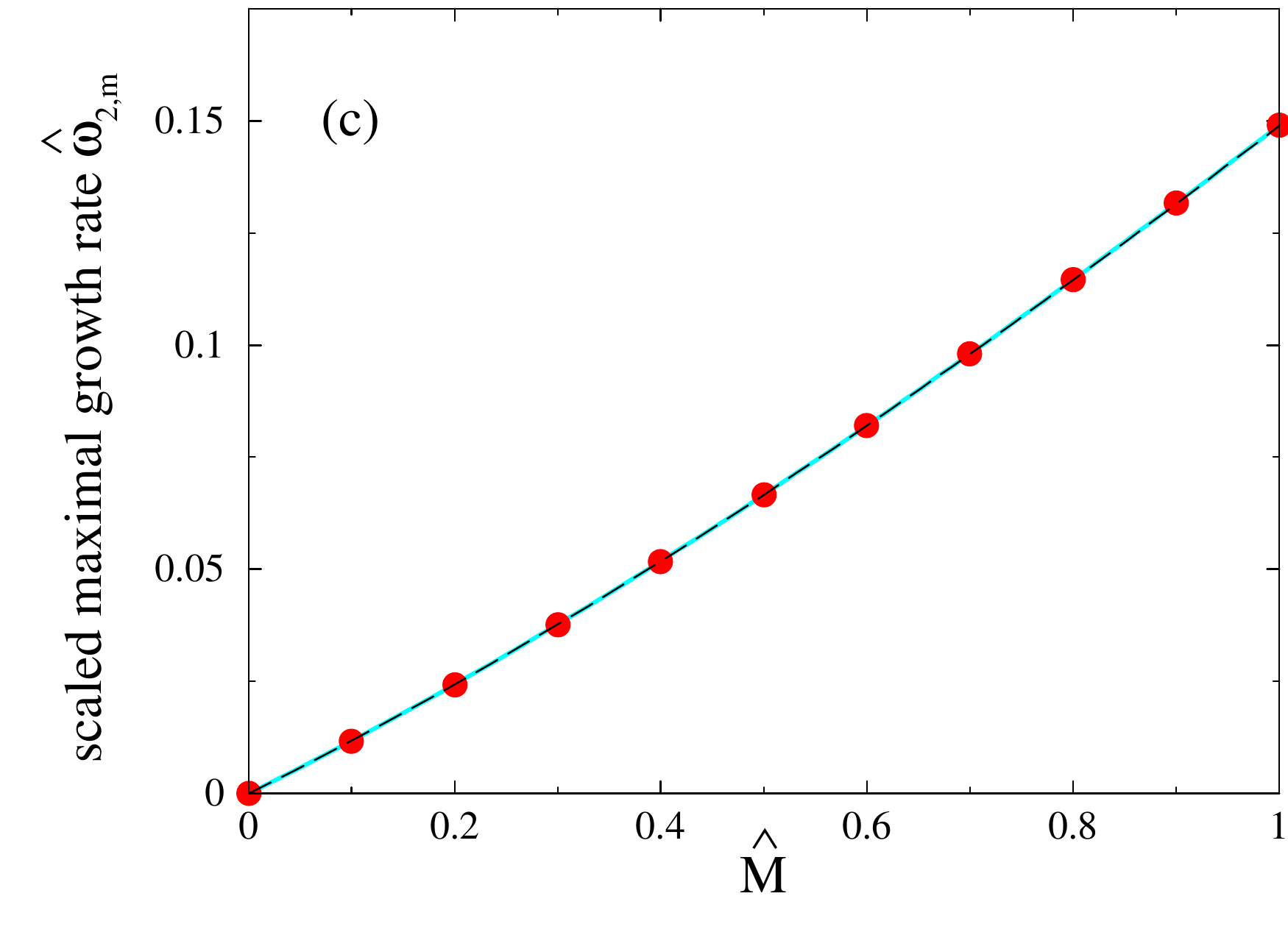} 
~\hskip 0.4 cm
\includegraphics[width=7.9cm]{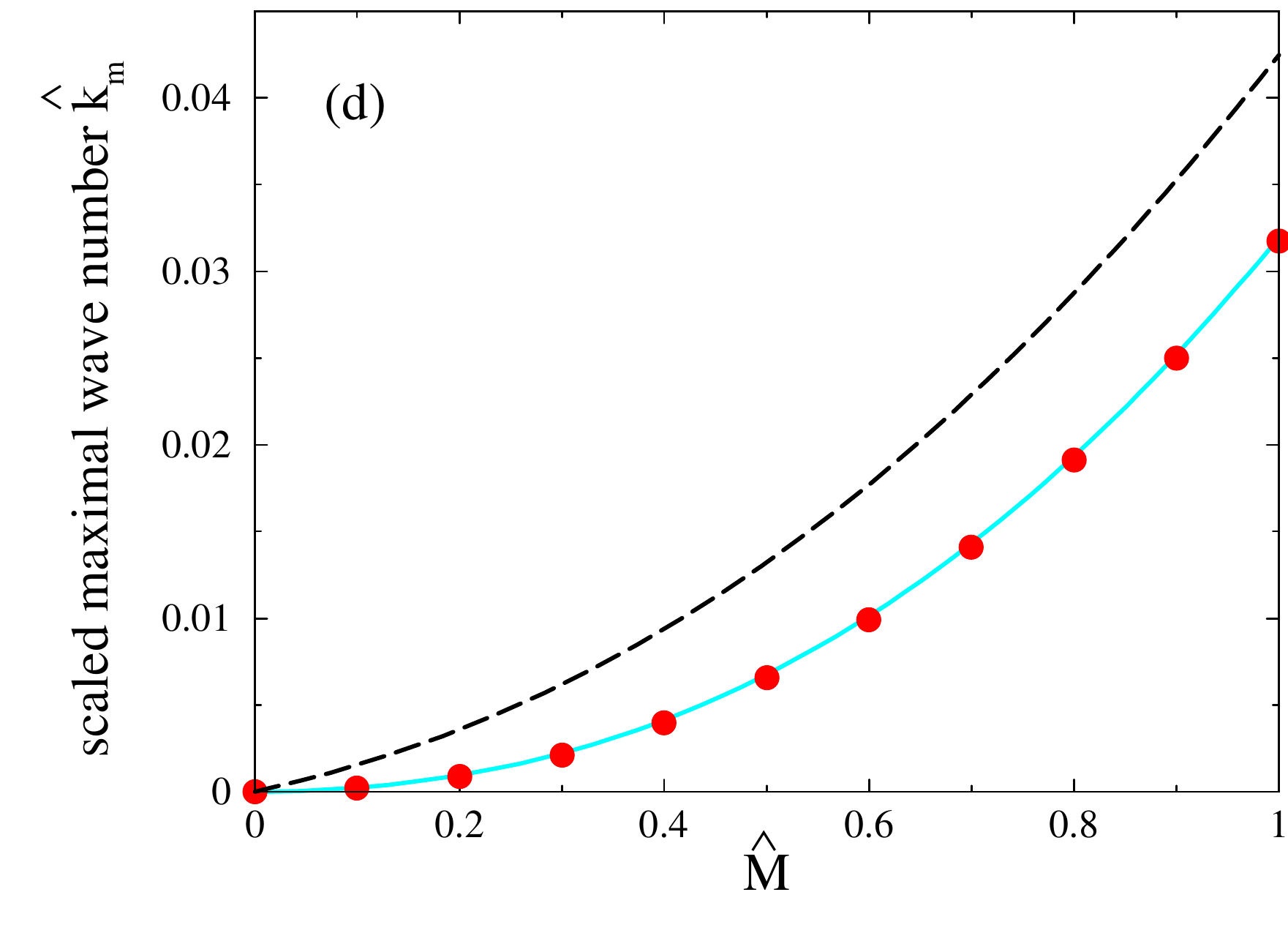} 
\caption{(Color online)
  The scaled maximal growth rate $\hat\omega_\mathrm{2,\,m}$ (a, c) and the scaled maximal wave number   $\hat k_\mathrm{m}$ (b, d) against the rescaled magnetization  for a magnetic fluid like   APG\,E32 but with a tenth of the original dynamical viscosity (a,b) and for APG\,E32 itself (c,d).
  The solid cyan (long-dashed black) line indicates the numerical data of the dispersion relation~(\ref{eq:disprel}) for $h=15$ ($9$) mm, the filled red circles the results of the scaling laws. The thin vertical dashed black lines in (a, b)
  display $\bar\nu^2/6$,  the limit of the validity of the scaling laws~(\ref{eq:hat_omega2_hatM})
  and~(\ref{eq:hat_k_hatM}).}
\label{fig:growthrate_wavenumber_vs_scaling_law}
\end{center}
\end{figure*}

It is known from a previous study for less viscous fluids \cite{lange01_wave} that a layer thickness of about the critical wavelength $\lambda_\mathrm{c}$ is necessary to represent the case $h\rightarrow\infty$ for the maximal growth rate as well as for the corresponding wave number, as shown in Fig.\,\ref{fig:growthrate_wavenumber_vs_scaling_law}(a, b).

That rule is no longer valid for more viscous fluids like APG\,E32 ($\lambda_\mathrm{c} \simeq 11$\,mm) as Fig.\ \ref{fig:growthrate_wavenumber_vs_scaling_law}(d) displays.
The results for the wave number deviate considerably from the results of the scaling law - compare the long-dashed black line ($h=9$\,mm) and the red filled circles in Fig.\ref{fig:growthrate_wavenumber_vs_scaling_law}(d).
By choosing a layer thickness of $h=15$ mm, the results stemming from the numerical solution of
the dispersion relation~(\ref{eq:disprel}) agree rather well with the data from the scaling laws - compare solid cyan lines and filled red circles in Fig.\ref{fig:growthrate_wavenumber_vs_scaling_law}(d).
Note that the maximal growth rate is not sensitive to $h$, as shown in Fig.
\ref{fig:growthrate_wavenumber_vs_scaling_law}(c).
As a r{\'e}sum{\'e} the rule can be formulated that for magnetic fluids with high viscosities a  larger filling depth than in the case of low viscosities has to be used,
in order to approximate the results of $h\rightarrow\infty$.

\section{\label{sec:results}Results and discussion}
We will next compare the experimentally determined growth rates, with the calculated ones for our particular fluid (Sect.\,\ref{subsec:growth}). Then we widen our scope and compare as well the decay rates with the model (Sect.\,\ref{subsec:decay}). Eventually some deviations are discussed in the context of structured ferrofluids (Sect.\,\ref{subsec:complex}).

\subsection{\label{subsec:growth} Comparing the growth rate in experiment and theory}
For our kind of magnetic fluids it was argued in the introduction that their high viscosity paves the way into a scaling regime, hitherto not accessible. That claim is now proven since a value of $87.9$ for the upper bound $\bar\nu^2/6$ of the scaling regime results, as summarized in Tab.\ \ref{tab:ingredients_growthrate_surfacetension}.
That means that for experimentally feasible scaled supercritical
magnetizations $\hat M$ the region $\hat M \leq \bar\nu^2/6$ is approachable.
The corresponding Eq.\ (\ref{eq:hat_omega2_hatM}) for the maximal growth rate states that $\hat\omega_\mathrm{2, m}$ should increase mainly linearly with $\hat M$ as long as $\hat M$ is not too large.

\begin{figure}[tbp]
\begin{center}
\includegraphics[width=7.9cm]{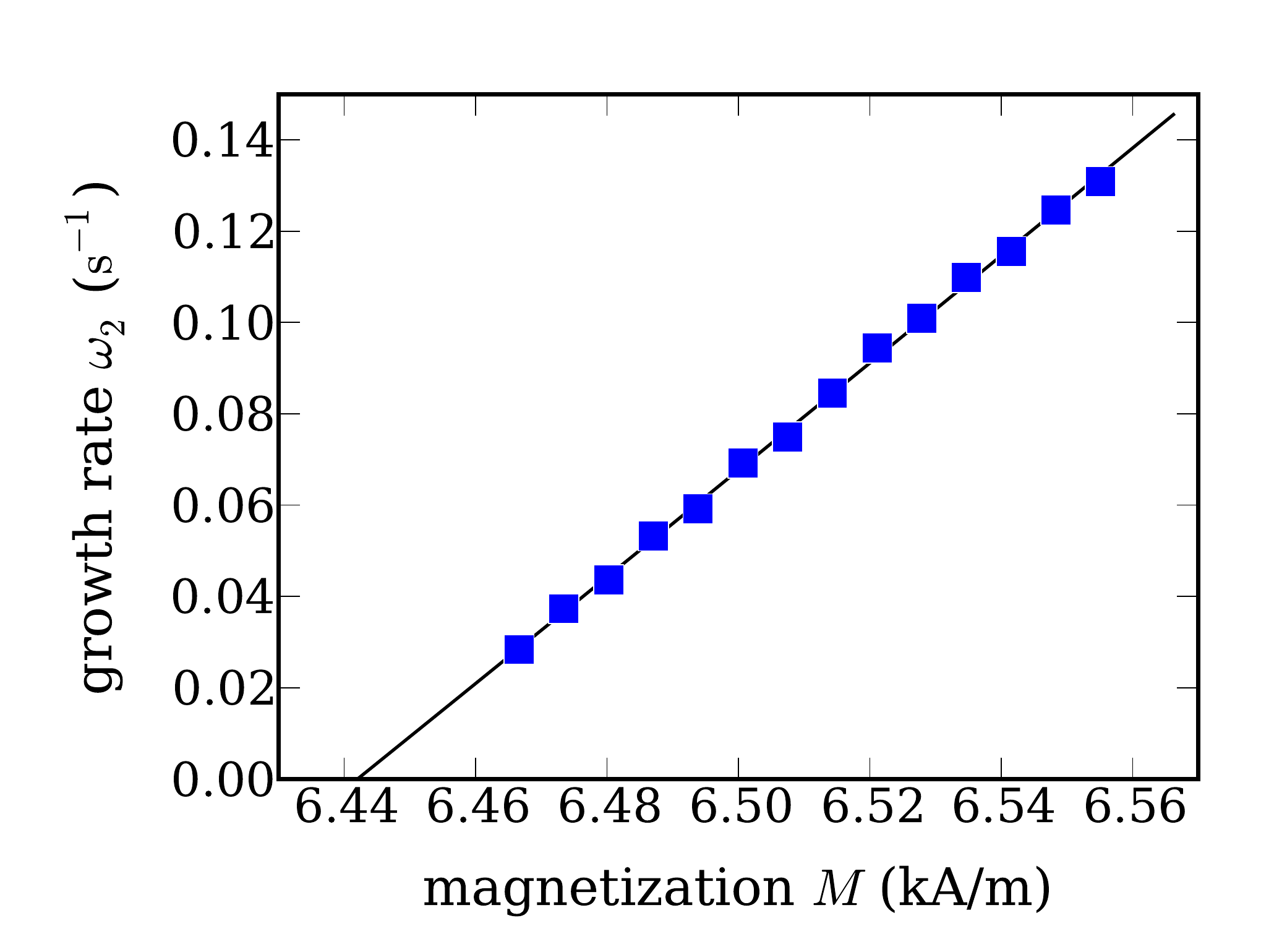}  
  \caption{(Color online) The linear growth rates $\omega_2$ ($\blacksquare$) from the experiment and the maximal growth rate $\omega_\mathrm{2, m}$ (solid black line) from theory as a function of the magnetization $M$.}
\label{fig:comparison_scaling_exp}
\end{center}
\end{figure}

To confirm this scaling behavior, the experimentally determined growth rates from
Fig.\,\ref{fig:growthrate_measurement} are plotted together with a Levenberg-Marquard fit \cite{press2002numerical} of the maximal growth rate obtained from  Eq.(\ref{eq:hat_omega2_hatM}) versus the the magnetisation $M$ as shown in Fig.\ \ref{fig:comparison_scaling_exp}. The agreement between both data sets is convincing. In table \ref{tab:fit.para} we present in line one the parameters for viscosity and surface tension, obtained from the fit. For comparison, we reprint in line zero the measured values. The fitted surface tension is well within the error bars of the measured value, whereas the fitted viscosity is only 6 \% below the measured one.
Thus the theoretically predicted linear dependence of the growth rates on $M$ is experimentally confirmed.

\begin{table*}[btp]
\caption{Results obtained for fitting Eq.\,(\ref{eq:hat_omega2_hatM}) to the experimental data. The filling depth was $h=5$\,mm, and the wavenumber of maximal growth was fixed to $k=544.44\,\mathrm{m^{-1}}$, as determined from the experiment, see Fig.\,\ref{fig:wachstum_fourier}. The symbol $\upharpoonleft$ marks a fit of growth data only, whereas  $\upharpoonleft\downharpoonright$ indicates that growth and decay data were taken into account.}
\begin{ruledtabular}
\begin{tabular}{lclllllll}
& direction & property  & $\eta_\mathrm{S}$ (Pa\,s) & $\vartheta$ $\left(\mathrm{\frac{Pa\,s^2}{rad}}\right)$ & $\sigma_\mathrm{S}$ $\left(\mathrm{\frac{mN}{m}}\right)$ & $\varsigma$ $\left(\mathrm{\frac{mN}{ms}}\right)$ & $M_\mathrm{c} \left(\mathrm{\frac{kA}{m}}\right)$ & $B_\mathrm{c}\left(\mathrm{mT}\right) $  \\ \hline \hline
0. & - & measured parameters   & $4.48 \pm 0.2$ & - & $30.9 \pm 5$ & - & - & - \\ \hline \hline
1. & $\upharpoonleft$                  & all static parameters   & 4.2 & 0 & 34.3 & 0  & 6.442 & 11.24\\ \hline
2. & $\upharpoonleft\downharpoonright$ & all static parameters   & 5.5 & 0 & 33.8 & 0  & 6.421 & 11.20\\ \hline
3. & $\upharpoonleft\downharpoonright$ & dynamic surface tension & 5.5 & 0 & 33.8 & -9.9 & 6.421 & 11.20 \\ \hline
4. & $\upharpoonleft\downharpoonright$ & non-Newtonian viscosity & 5.2 & -4.8 & 34.1 & 0  & 6.434 & 11.22 \\
\end{tabular}
\end{ruledtabular}
\label{tab:fit.para}
\end{table*}

\subsection{\label{subsec:decay}Comparing growth and decay}
\begin{figure}[tbp]
\begin{center}
\includegraphics[width=7.9cm]{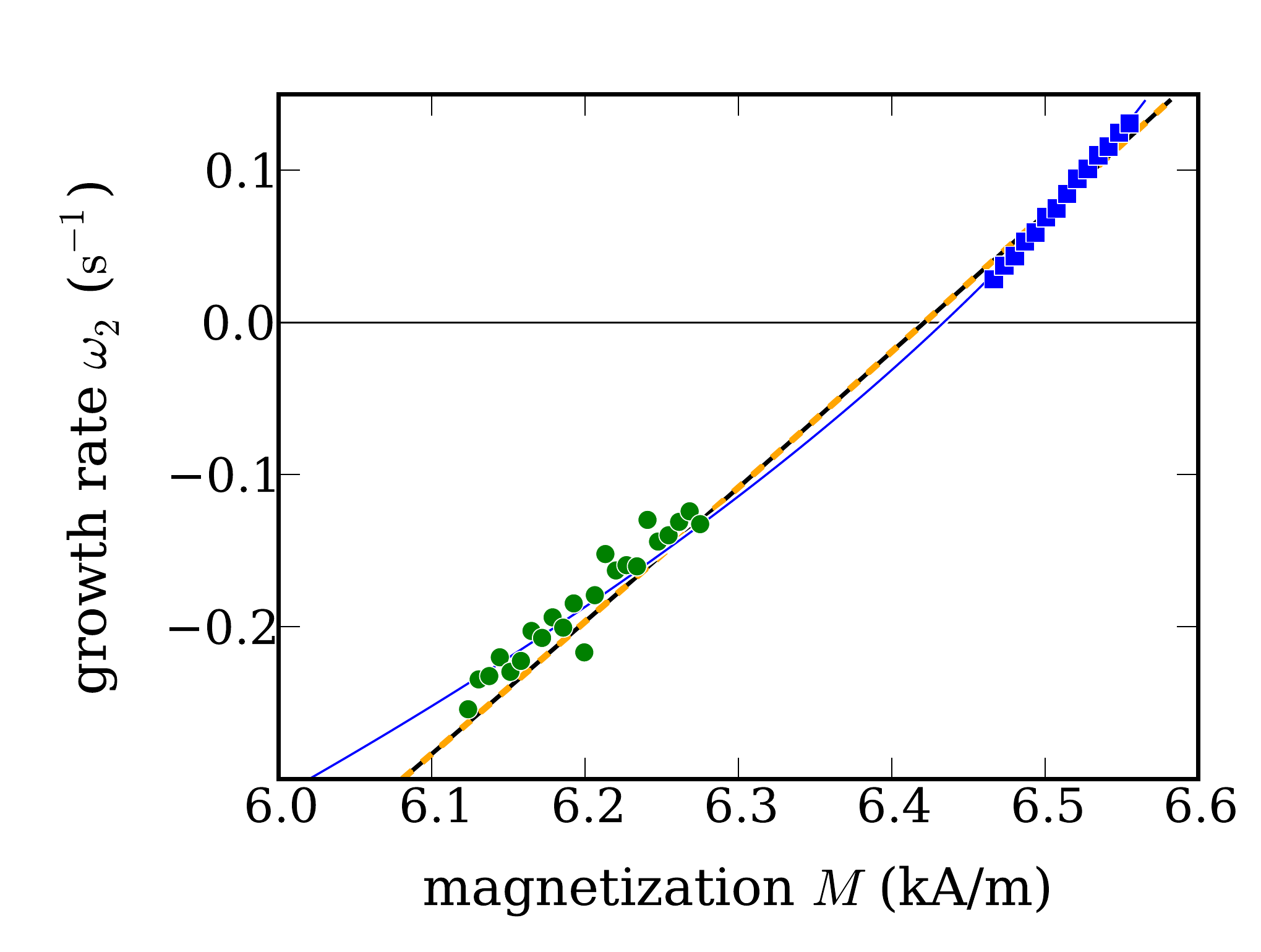}  
~\hskip 0.4 cm
\caption{(Color online) The growth ($\blacksquare$) and decay ($\bullet$)rates $\omega_2$, respectively, of the pattern amplitude as a function of the magnetization $M$. The symbols represent the measured data.
The black dashed line shows a fit of Eq.\,(\ref{eq:hat_omega2_hatM}) to the experimental growth and decay rates, with the parameters given in the line two of table \ref{tab:fit.para}. The orange dashed line marks as well a fit by Eq.\,(\ref{eq:hat_omega2_hatM}), but is taking into account a growth rate dependent surface tension, as described by  Eq.\,(\ref{eq:dyn.sigma}). For the parameters see line three of table \ref{tab:fit.para}.
The solid blue line displays a fit taking into account a growth-rate dependent viscosity according to Eq. (\ref{eq:dyn.eta}). For parameters see line four of table \ref{tab:fit.para}.}
\label{fig:growthrate_comparison}
\end{center}
\end{figure}

Next we focus as well on the experimental data for the decay, which are plotted together with the growth data in Fig.\ \ref{fig:growthrate_comparison}. The decay rates ($ \bullet $) are scattering more widely in comparison to the growth rates ($\blacksquare$). This may be due to the fact, that the decay rates could not be resolved in the bistability range, and thus not in the immediate vicinity of $M_\mathrm{c}$, in contrast to the growth rates. The black dashed line marks the outcome of a fit of Eq\,(\ref{eq:hat_omega2_hatM}) to \emph{all} experimental values. Also in this extended range the fit describes the measured growth and decay rates to some extent. In table \ref{tab:fit.para} we present in line two the fit parameters for viscosity and surface tension. The fitted surface tension is well within the error bars of the measured value, whereas the fitted viscosity is about 20 \% above the measured one.

Most importantly, inspecting the measured data more closely, one observes a different inclination for growth and decay rates with respect to $M$. Obviously this systematic deviation is not matched by Eq.\,(\ref{eq:hat_omega2_hatM}). As a possible origin for the different inclinations one may suspect that utilizing the static surface tension in Eq.(\ref{eq:hat_omega2_hatM}) is not a sufficient approximation.
Indeed during the growth of the peaks new surface area is generated, and the diffusion of surfactants from the bulk of the ferrofluid towards the surface may lag behind. Similarly during the decay of the peaks surface area is annihilated and the surface density of surfactants may there exceed the equilibrium concentration. Therefore we adopt a growth-rate-dependent dynamic surface tension according to
\begin{equation}
\sigma(\omega_2) = \sigma_\mathrm{S} + \omega_2 \cdot \varsigma,
\label{eq:dyn.sigma}
\end{equation}
where $\sigma_\mathrm{S}$ denotes the static surface tension and $\varsigma$ a coefficient of dimension $\mathrm{N\cdot(m~s)^{-1}}$.
In Fig.\,\ref{fig:growthrate_comparison} the orange dashed line marks the outcome of the fit.
It follows the black dashed line, and thus can not explain the different inclinations.

In a next attempt to describe the different inclinations we postulate a growth rate dependent viscosity in the form of
\begin{equation}
\eta (\omega) = \eta_\mathrm{S} + \omega_2 \cdot \vartheta,
\label{eq:dyn.eta}
\end{equation}
where $\vartheta$ is a coefficient of dimension $\mathrm{Pa\,s^2}$.
In Fig.\,\ref{fig:growthrate_comparison} a fit by Eq.\,(\ref{eq:dyn.eta}) is marked by the solid blue line. Obviously this phenomenological ansatz meets the data remarkeably well.

\subsection{Discussion of deviations} \label{subsec:complex}
A possible explanation of this complex behaviour is based upon the formation of chains of magnetic particles, which is indicated by the enhanced shear thinning as recorded in Fig.\,\ref{fig:shear.thinning}.
The chain formation will be most prominent in the higher magnetic field in the spikes at the starting amplitude $A_\mathrm{high}$, marked in Fig.\,\ref{fig:frozenprotocol}. These chains are then increasing the magnetoviscosity during the decay of the spikes, which retards the decay (cf.\,path $\downarrow$\textit{2} and $\downarrow$\textit{3b} in Fig.\ref{fig:frozenprotocol}). During the decay they are partially destroyed. As a consequence, after switching again to an overcritical induction, the growth of the spikes (path $\uparrow$\textit{3a}) is comparatively faster. In contrast, our theory is based on Newtonian fluids. An extension to shear thinning and structured liquids has still to be developed.

We are next comparing the critical inductions in the last column of table \ref{tab:fit.para}. The  static fit of the growth process yields $B_\mathrm{c1}=11.24\,\mathrm{mT}$ and deviates by only 1\% from the mean value $\bar{B_\mathrm{c}}=11.36\,\mathrm{mT}$ obtained by a fit of the full dynamics by means of amplitude equations in Ref.\,\cite{gollwitzer10}. All other values for $B_\mathrm{c}$ underestimate this value slightly more (cf.\,line 2-4). In the latter three cases the growth \emph{and} decay was taken into account. This is a conformation, that mainly  the decay is affected by chain formation in the spikes.

Eventually we will not hide \emph{four} further effects which may have impact on our experiment:

\emph{First}, the experiments are performed in a finite container which comprises only 27 spikes on a hexagonal lattice, whereas the theory considers a laterally infinite layer.
Our finite circular size does indeed suppress the onset of a hexagonal pattern,
due to the ramp described above.

\emph{Second}, by seeding a regular hexagonal pattern at large amplitude the selected wavelength may differ from the wavelength of maximal growth. This can in principle shift the experimental threshold towards higher values. However, it was demonstrated by linear stability analysis that this effect can be neglected in the limit of high viscosities \cite{reimann03}.

\emph{Third}, magnetophoresis may take place in the crests of the pattern, in this way creating an inhomogeneous distribution of magnetite. Even though the timescale for separation in a low viscous MF comprise days \cite{ivanov10,lavrova10} and our measurements last only hours, an effect can not completely excluded.

A \emph{fourth} reason may be that instead of the shear viscosity the extensional viscosity has to be taken into account in Eq.(\ref{eq:disprel}). Indeed, besides a small viscous sublayer, the flow profile of surface waves can "be described by a potential and is rotational free and purely elongational"  \cite{kityk2006}. Most recently a capillary-break-up-extensional-rheometer was subjected to magnetic fields oriented along the direction of the capillary \cite{galindo2015}. For increasing fields an enlarged elongational viscosity was observed. This effect was also attributed to chain formation. However, to measure the elongational viscosity of ferrofluids is a difficult task, and sensitive devices have still to be developed.

\section{\label{sec:conclusion}Conclusion}
Using a highly viscous magnetic fluid, the dynamics of the formation of the
Rosensweig instability can be slowed down to the order of minutes. Therefore, it
is possible to measure the dynamics using a two-dimensional imaging technique,
in contrast to previous work \cite{knieling07}, where only a one-dimensional cut through the two-dimensional pattern was accomplished. By means of a specific measurement protocol we were able to seed regular patterns of small amplitude, suitable for a comparison with linear theory. From the evolution of their amplitudes we could estimate the linear growth and decay rates, respectively.
Our experiment confirmed for the very first time a \emph{linear} scaling of the growth rate with the magnetic inductions, as predicted \cite{lange01_growth} for the immediate vicinity of the bifurcation point. Thus the scaling behavior of the growth rate is now confirmed for supercritical magnetizations not only above \cite{knieling07} but also below the boundary of the two scaling regimes at $\bar\nu^2/6$.

Additionally, we uncovered, that the rates of growth and decay are slightly different, a phenomenon not predicted by the theory. A possible origin of this discrepancy is the formation of chains of magnetic particles. Their presence in our ferrofluid is indicated by the magnetically enhanced shear thinning. The build up of chains in the static spikes, and their subsequent destruction during the decay may change the effective viscosity of the structured ferrofluid, and thus explain the deviations.

So far our theory is based on Newtonian liquids. An extension to shear thinning and structured ferrofluids is referred to future investigations. It may be able to reproduce the scaling of the effective viscosity as described phenomenologically by Eq.\,({\ref{eq:scaling_visc})}.

\section*{Acknowledgements}
We thank M. M\"arkl for measuring the surface tension of the used magnetic fluid. The
temperature-controlled container was made with the help of Klaus Oetter and the mechanical
and electronic workshop the university of Bayreuth. Moreover discussions with Thomas Friedrich, Werner K\"ohler, Konstantin Morozov and Christian Wagner are gratefully acknowledged. R.R. is deeply indebted to the Emil-Warburg foundation for financially supporting repair and upgrade of the magnetorheometer.

\appendix
\section{\label{sec:appendix_coeff} }
The coefficients for the forth and fifth order of $\hat M$ in the scaling
laws~(\ref{eq:hat_omega2_hatM}, \ref{eq:hat_k_hatM}) are
\begin{widetext}
\begin{align}
\nonumber
  \Theta =& \frac{b_\chi}{\bar\nu} - \frac{6a_\chi(4+3a_\chi)+3b_\chi(8a_\chi+b_\chi+10)+3}{4\bar\nu^3}\\
\label{eq:theta}
         & + \frac{3(2+a_\chi)^2(3a_\chi^2+32a_\chi+22+10b_\chi)}{\bar\nu^5} -\frac{85(2+a_\chi)^4}{32\bar\nu^7}\; ,\\
\nonumber
  \xi =& \frac{6a_\chi(4+3a_\chi)+3b_\chi(8a_\chi+b_\chi+10)+3}{2\bar\nu^2}
          - \frac{3(2+a_\chi)^2(9a_\chi^2+80a_\chi+58+22b_\chi)}{8\bar\nu^4}\\
\label{eq:xi}
         & +\frac{99(2+a_\chi)^4}{\bar\nu^6}\; ,\\
\nonumber
  \iota =& - \frac{3a_\chi(2a_\chi+1)+6b_\chi(3a_\chi+2+b_\chi)}{2\bar\nu^3}
            - \frac{(2+a_\chi)^3(93a_\chi^2+712a_\chi+170b_\chi+542)}{16\bar\nu^7}\\
\nonumber
        & + \frac{3(2+a_\chi)\left[12a_\chi^3+a_\chi^2(6b_\chi+79)+a_\chi(54b_\chi+102)+b_\chi(54+5b_\chi)+29\right]}{4\bar\nu^5}\\
\label{eq:iota}
        & + \frac{407(2+a_\chi)^5}{64\bar\nu^9}\; ,\\
\nonumber
  o  =&  \frac{3a_\chi(2a_\chi+1)+6b_\chi(3a_\chi+2+b_\chi)}{\bar\nu^2}
         +\frac{3(2+a_\chi)^3(23a_\chi^2+158a_\chi+33b_\chi+125)}{\bar\nu^6}\\
\nonumber
        & - \frac{3(2+a_\chi)\left[36a_\chi^3+a_\chi^2(18b_\chi+217)+a_\chi(138b_\chi+282)+b_\chi(138+11b_\chi)+83\right]}{4\bar\nu^4}\\
\label{eq:o}
        & - \frac{491(2+a_\chi)^5}{32\bar\nu^8} \; .
\end{align}
\end{widetext}

\begin{thebibliography}{45}%
\makeatletter
\providecommand \@ifxundefined [1]{%
 \@ifx{#1\undefined}
}%
\providecommand \@ifnum [1]{%
 \ifnum #1\expandafter \@firstoftwo
 \else \expandafter \@secondoftwo
 \fi
}%
\providecommand \@ifx [1]{%
 \ifx #1\expandafter \@firstoftwo
 \else \expandafter \@secondoftwo
 \fi
}%
\providecommand \natexlab [1]{#1}%
\providecommand \enquote  [1]{``#1''}%
\providecommand \bibnamefont  [1]{#1}%
\providecommand \bibfnamefont [1]{#1}%
\providecommand \citenamefont [1]{#1}%
\providecommand \href@noop [0]{\@secondoftwo}%
\providecommand \href [0]{\begingroup \@sanitize@url \@href}%
\providecommand \@href[1]{\@@startlink{#1}\@@href}%
\providecommand \@@href[1]{\endgroup#1\@@endlink}%
\providecommand \@sanitize@url [0]{\catcode `\\12\catcode `\$12\catcode
  `\&12\catcode `\#12\catcode `\^12\catcode `\_12\catcode `\%12\relax}%
\providecommand \@@startlink[1]{}%
\providecommand \@@endlink[0]{}%
\providecommand \url  [0]{\begingroup\@sanitize@url \@url }%
\providecommand \@url [1]{\endgroup\@href {#1}{\urlprefix }}%
\providecommand \urlprefix  [0]{URL }%
\providecommand \Eprint [0]{\href }%
\providecommand \doibase [0]{http://dx.doi.org/}%
\providecommand \selectlanguage [0]{\@gobble}%
\providecommand \bibinfo  [0]{\@secondoftwo}%
\providecommand \bibfield  [0]{\@secondoftwo}%
\providecommand \translation [1]{[#1]}%
\providecommand \BibitemOpen [0]{}%
\providecommand \bibitemStop [0]{}%
\providecommand \bibitemNoStop [0]{.\EOS\space}%
\providecommand \EOS [0]{\spacefactor3000\relax}%
\providecommand \BibitemShut  [1]{\csname bibitem#1\endcsname}%
\let\auto@bib@innerbib\@empty
\bibitem [{\citenamefont {Edgeworth}\ \emph {et~al.}(1984)\citenamefont
  {Edgeworth}, \citenamefont {Dalton},\ and\ \citenamefont
  {Parnell}}]{edgeworth84}%
  \BibitemOpen
  \bibfield  {author} {\bibinfo {author} {\bibfnamefont {R.}~\bibnamefont
  {Edgeworth}}, \bibinfo {author} {\bibfnamefont {B.~J.}\ \bibnamefont
  {Dalton}}, \ and\ \bibinfo {author} {\bibfnamefont {T.}~\bibnamefont
  {Parnell}},\ }\href@noop {} {\bibfield  {journal} {\bibinfo  {journal} {Eur.
  J. Phys.}\ }\textbf {\bibinfo {volume} {5}},\ \bibinfo {pages} {198}
  (\bibinfo {year} {1984})}\BibitemShut {NoStop}%
\bibitem [{\citenamefont {Eggers}(1997)}]{eggers97}%
  \BibitemOpen
  \bibfield  {author} {\bibinfo {author} {\bibfnamefont {J.}~\bibnamefont
  {Eggers}},\ }\href@noop {} {\bibfield  {journal} {\bibinfo  {journal} {Rev.
  Mod. Phys.}\ }\textbf {\bibinfo {volume} {69}},\ \bibinfo {pages} {865}
  (\bibinfo {year} {1997})}\BibitemShut {NoStop}%
\bibitem [{\citenamefont {Rothert}\ \emph {et~al.}(2001)\citenamefont
  {Rothert}, \citenamefont {Richter},\ and\ \citenamefont
  {Rehberg}}]{rothert01}%
  \BibitemOpen
  \bibfield  {author} {\bibinfo {author} {\bibfnamefont {A.}~\bibnamefont
  {Rothert}}, \bibinfo {author} {\bibfnamefont {R.}~\bibnamefont {Richter}}, \
  and\ \bibinfo {author} {\bibfnamefont {I.}~\bibnamefont {Rehberg}},\
  }\href@noop {} {\bibfield  {journal} {\bibinfo  {journal} {Phys. Rev. Lett.}\
  }\textbf {\bibinfo {volume} {87}},\ \bibinfo {pages} {084501} (\bibinfo
  {year} {2001})}\BibitemShut {NoStop}%
\bibitem [{\citenamefont {of~Queensland}()}]{queensland2013}%
  \BibitemOpen
  \bibfield  {author} {\bibinfo {author} {\bibfnamefont {U.}~\bibnamefont
  {of~Queensland}},\ }\href@noop {} {\enquote {\bibinfo {title} {The pitch drop
  experiment},}\ }\bibinfo {note}
  {{http://smp.uq.edu.au/content/pitch-drop-experiment (download
  26.04.2013)}}\BibitemShut {NoStop}%
\bibitem [{the()}]{thetenth}%
  \BibitemOpen
  \href@noop {} {}\bibinfo {howpublished} {www.thetenthwatch.com (download
  13.2.2015)}\BibitemShut {NoStop}%
\bibitem [{\citenamefont {Richter}(2011)}]{richter11}%
  \BibitemOpen
  \bibfield  {author} {\bibinfo {author} {\bibfnamefont {R.}~\bibnamefont
  {Richter}},\ }\href@noop {} {\bibfield  {journal} {\bibinfo  {journal}
  {Europhys. News}\ }\textbf {\bibinfo {volume} {42}},\ \bibinfo {pages} {17}
  (\bibinfo {year} {2011})}\BibitemShut {NoStop}%
\bibitem [{\citenamefont {Castellvecchi}(2005)}]{castellvecchi05}%
  \BibitemOpen
  \bibfield  {author} {\bibinfo {author} {\bibfnamefont {D.}~\bibnamefont
  {Castellvecchi}},\ }\href@noop {} {\enquote {\bibinfo {title} {New instrument
  for solo performance},}\ }\bibinfo {howpublished} {Physical Revie Focus}
  (\bibinfo {year} {http://focus.aps.org/story/v15/st18 (2005)})\BibitemShut
  {NoStop}%
\bibitem [{\citenamefont {Cowley}\ and\ \citenamefont
  {Rosensweig}(1967)}]{cowley67}%
  \BibitemOpen
  \bibfield  {author} {\bibinfo {author} {\bibfnamefont {M.~D.}\ \bibnamefont
  {Cowley}}\ and\ \bibinfo {author} {\bibfnamefont {R.~E.}\ \bibnamefont
  {Rosensweig}},\ }\href@noop {} {\bibfield  {journal} {\bibinfo  {journal} {J.
  Fluid Mech.}\ }\textbf {\bibinfo {volume} {30}},\ \bibinfo {pages} {671}
  (\bibinfo {year} {1967})}\BibitemShut {NoStop}%
\bibitem [{\citenamefont {Rosensweig}(1985{\natexlab{a}})}]{rosensweig85_book}%
  \BibitemOpen
  \bibfield  {author} {\bibinfo {author} {\bibfnamefont {R.~E.}\ \bibnamefont
  {Rosensweig}},\ }\href@noop {} {\emph {\bibinfo {title}
  {Ferrohydrodynamics}}}\ (\bibinfo  {publisher} {Cambridge University Press},\
  \bibinfo {address} {Cambridge},\ \bibinfo {year} {1985})\BibitemShut
  {NoStop}%
\bibitem [{\citenamefont {Salin}(1993)}]{salin93}%
  \BibitemOpen
  \bibfield  {author} {\bibinfo {author} {\bibfnamefont {D.}~\bibnamefont
  {Salin}},\ }\href@noop {} {\bibfield  {journal} {\bibinfo  {journal}
  {Europhys. Lett.}\ }\textbf {\bibinfo {volume} {21}},\ \bibinfo {pages} {667}
  (\bibinfo {year} {1993})}\BibitemShut {NoStop}%
\bibitem [{\citenamefont {Weilepp}\ and\ \citenamefont
  {Brand}(1996)}]{weilepp96}%
  \BibitemOpen
  \bibfield  {author} {\bibinfo {author} {\bibfnamefont {J.}~\bibnamefont
  {Weilepp}}\ and\ \bibinfo {author} {\bibfnamefont {H.~R.}\ \bibnamefont
  {Brand}},\ }\href@noop {} {\bibfield  {journal} {\bibinfo  {journal} {J.
  Phys. II France}\ }\textbf {\bibinfo {volume} {6}},\ \bibinfo {pages} {419}
  (\bibinfo {year} {1996})}\BibitemShut {NoStop}%
\bibitem [{\citenamefont {Lange}\ \emph {et~al.}(2000)\citenamefont {Lange},
  \citenamefont {Reimann},\ and\ \citenamefont {Richter}}]{lange00_wave}%
  \BibitemOpen
  \bibfield  {author} {\bibinfo {author} {\bibfnamefont {A.}~\bibnamefont
  {Lange}}, \bibinfo {author} {\bibfnamefont {B.}~\bibnamefont {Reimann}}, \
  and\ \bibinfo {author} {\bibfnamefont {R.}~\bibnamefont {Richter}},\
  }\href@noop {} {\bibfield  {journal} {\bibinfo  {journal} {Phys. Rev. E}\
  }\textbf {\bibinfo {volume} {61}},\ \bibinfo {pages} {5528} (\bibinfo {year}
  {2000})}\BibitemShut {NoStop}%
\bibitem [{\citenamefont {Lange}\ \emph {et~al.}(2001)\citenamefont {Lange},
  \citenamefont {Reimann},\ and\ \citenamefont {Richter}}]{lange01_wave}%
  \BibitemOpen
  \bibfield  {author} {\bibinfo {author} {\bibfnamefont {A.}~\bibnamefont
  {Lange}}, \bibinfo {author} {\bibfnamefont {B.}~\bibnamefont {Reimann}}, \
  and\ \bibinfo {author} {\bibfnamefont {R.}~\bibnamefont {Richter}},\
  }\href@noop {} {\bibfield  {journal} {\bibinfo  {journal}
  {Magnetohydrodynamics}\ }\textbf {\bibinfo {volume} {37}},\ \bibinfo {pages}
  {261} (\bibinfo {year} {2001})}\BibitemShut {NoStop}%
\bibitem [{\citenamefont {Knieling}\ \emph {et~al.}(2007)\citenamefont
  {Knieling}, \citenamefont {Richter}, \citenamefont {Rehberg}, \citenamefont
  {Matthies},\ and\ \citenamefont {Lange}}]{knieling07}%
  \BibitemOpen
  \bibfield  {author} {\bibinfo {author} {\bibfnamefont {H.}~\bibnamefont
  {Knieling}}, \bibinfo {author} {\bibfnamefont {R.}~\bibnamefont {Richter}},
  \bibinfo {author} {\bibfnamefont {I.}~\bibnamefont {Rehberg}}, \bibinfo
  {author} {\bibfnamefont {G.}~\bibnamefont {Matthies}}, \ and\ \bibinfo
  {author} {\bibfnamefont {A.}~\bibnamefont {Lange}},\ }\href@noop {}
  {\bibfield  {journal} {\bibinfo  {journal} {Phys. Rev. E}\ }\textbf {\bibinfo
  {volume} {76}},\ \bibinfo {pages} {066301} (\bibinfo {year}
  {2007})}\BibitemShut {NoStop}%
\bibitem [{\citenamefont {Gollwitzer}\ \emph {et~al.}()\citenamefont
  {Gollwitzer}, \citenamefont {Rehberg}, \citenamefont {Lange},\ and\
  \citenamefont {Richter}}]{gollwitzer2009frozensweig}%
  \BibitemOpen
  \bibfield  {author} {\bibinfo {author} {\bibfnamefont {C.}~\bibnamefont
  {Gollwitzer}}, \bibinfo {author} {\bibfnamefont {I.}~\bibnamefont {Rehberg}},
  \bibinfo {author} {\bibfnamefont {A.}~\bibnamefont {Lange}}, \ and\ \bibinfo
  {author} {\bibfnamefont {R.}~\bibnamefont {Richter}},\ }\href@noop {}
  {\enquote {\bibinfo {title} {''{Frozensweig}'': a cool instability in the
  limit $\eta\rightarrow\infty$},}\ }\bibinfo {note} {Book of abstracts of the
  9th German Ferrofluid Workshop, Benediktbeuern (2009)}\BibitemShut {NoStop}%
\bibitem [{\citenamefont {Richter}\ and\ \citenamefont
  {Bl\"asing}(2001)}]{richter01}%
  \BibitemOpen
  \bibfield  {author} {\bibinfo {author} {\bibfnamefont {R.}~\bibnamefont
  {Richter}}\ and\ \bibinfo {author} {\bibfnamefont {J.}~\bibnamefont
  {Bl\"asing}},\ }\href@noop {} {\bibfield  {journal} {\bibinfo  {journal}
  {Rev. Sci. Instrum.}\ }\textbf {\bibinfo {volume} {72}},\ \bibinfo {pages}
  {1729} (\bibinfo {year} {2001})}\BibitemShut {NoStop}%
\bibitem [{\citenamefont {Gollwitzer}\ \emph
  {et~al.}(2007{\natexlab{a}})\citenamefont {Gollwitzer}, \citenamefont
  {Matthies}, \citenamefont {Richter}, \citenamefont {Rehberg},\ and\
  \citenamefont {Tobiska}}]{gollwitzer07}%
  \BibitemOpen
  \bibfield  {author} {\bibinfo {author} {\bibfnamefont {C.}~\bibnamefont
  {Gollwitzer}}, \bibinfo {author} {\bibfnamefont {G.}~\bibnamefont
  {Matthies}}, \bibinfo {author} {\bibfnamefont {R.}~\bibnamefont {Richter}},
  \bibinfo {author} {\bibfnamefont {I.}~\bibnamefont {Rehberg}}, \ and\
  \bibinfo {author} {\bibfnamefont {L.}~\bibnamefont {Tobiska}},\ }\href@noop
  {} {\bibfield  {journal} {\bibinfo  {journal} {J. Fluid Mech.}\ }\textbf
  {\bibinfo {volume} {571}},\ \bibinfo {pages} {455} (\bibinfo {year}
  {2007}{\natexlab{a}})}\BibitemShut {NoStop}%
\bibitem [{\citenamefont {Gollwitzer}\ \emph {et~al.}(2010)\citenamefont
  {Gollwitzer}, \citenamefont {Rehberg},\ and\ \citenamefont
  {Richter}}]{gollwitzer10}%
  \BibitemOpen
  \bibfield  {author} {\bibinfo {author} {\bibfnamefont {C.}~\bibnamefont
  {Gollwitzer}}, \bibinfo {author} {\bibfnamefont {I.}~\bibnamefont {Rehberg}},
  \ and\ \bibinfo {author} {\bibfnamefont {R.}~\bibnamefont {Richter}},\
  }\href@noop {} {\bibfield  {journal} {\bibinfo  {journal} {New J. Phys.}\
  }\textbf {\bibinfo {volume} {12}},\ \bibinfo {pages} {093037} (\bibinfo
  {year} {2010})}\BibitemShut {NoStop}%
\bibitem [{\citenamefont {Lloyd}\ \emph {et~al.}(2015)\citenamefont {Lloyd},
  \citenamefont {Gollwitzer}, \citenamefont {Rehberg},\ and\ \citenamefont
  {Richter}}]{lloyd2015}%
  \BibitemOpen
  \bibfield  {author} {\bibinfo {author} {\bibfnamefont {D.~J.~B.}\
  \bibnamefont {Lloyd}}, \bibinfo {author} {\bibfnamefont {C.}~\bibnamefont
  {Gollwitzer}}, \bibinfo {author} {\bibfnamefont {I.}~\bibnamefont {Rehberg}},
  \ and\ \bibinfo {author} {\bibfnamefont {R.}~\bibnamefont {Richter}},\ }\href
  {\doibase 10.1017/jfm.2015.565} {\bibfield  {journal} {\bibinfo  {journal}
  {J. Fluid Mech.}\ }\textbf {\bibinfo {volume} {783}},\ \bibinfo
  {pages} {283} (\bibinfo {year} {2015})}\BibitemShut {NoStop}%
\bibitem [{\citenamefont {Lange}(2001)}]{lange01_growth}%
  \BibitemOpen
  \bibfield  {author} {\bibinfo {author} {\bibfnamefont {A.}~\bibnamefont
  {Lange}},\ }\href@noop {} {\bibfield  {journal} {\bibinfo  {journal}
  {Europhys. Lett.}\ }\textbf {\bibinfo {volume} {55}},\ \bibinfo {pages} {327}
  (\bibinfo {year} {2001})}\BibitemShut {NoStop}%
\bibitem [{\citenamefont {Knieling}\ \emph {et~al.}(2010)\citenamefont
  {Knieling}, \citenamefont {Rehberg},\ and\ \citenamefont
  {Richter}}]{knieling10}%
  \BibitemOpen
  \bibfield  {author} {\bibinfo {author} {\bibfnamefont {H.}~\bibnamefont
  {Knieling}}, \bibinfo {author} {\bibfnamefont {I.}~\bibnamefont {Rehberg}}, \
  and\ \bibinfo {author} {\bibfnamefont {R.}~\bibnamefont {Richter}},\ }in\
  \href@noop {} {\emph {\bibinfo {booktitle} {Physics Procedia}}},\
  Vol.~\bibinfo {volume} {9},\ \bibinfo {editor} {edited by\ \bibinfo {editor}
  {\bibfnamefont {H.}~\bibnamefont {Yamaguchi}}},\ \bibinfo {organization}
  {12th International Conference on Magnetic Fluids}\ (\bibinfo  {publisher}
  {Elsevier},\ \bibinfo {address} {Amsterdam},\ \bibinfo {year} {2010})\ pp.\
  \bibinfo {pages} {199--204}\BibitemShut {NoStop}%
\bibitem [{\citenamefont {Gollwitzer}\ \emph {et~al.}(2009)\citenamefont
  {Gollwitzer}, \citenamefont {Krekhova}, \citenamefont {Lattermann},
  \citenamefont {Rehberg},\ and\ \citenamefont {Richter}}]{gollwitzer09}%
  \BibitemOpen
  \bibfield  {author} {\bibinfo {author} {\bibfnamefont {C.}~\bibnamefont
  {Gollwitzer}}, \bibinfo {author} {\bibfnamefont {M.}~\bibnamefont
  {Krekhova}}, \bibinfo {author} {\bibfnamefont {G.}~\bibnamefont
  {Lattermann}}, \bibinfo {author} {\bibfnamefont {I.}~\bibnamefont {Rehberg}},
  \ and\ \bibinfo {author} {\bibfnamefont {R.}~\bibnamefont {Richter}},\
  }\href@noop {} {\bibfield  {journal} {\bibinfo  {journal} {Soft Matter}\
  }\textbf {\bibinfo {volume} {5}},\ \bibinfo {pages} {2093} (\bibinfo {year}
  {2009})}\BibitemShut {NoStop}%
\bibitem [{\citenamefont {Ivanov}\ and\ \citenamefont
  {Kuznetsova}(2001)}]{ivanov01}%
  \BibitemOpen
  \bibfield  {author} {\bibinfo {author} {\bibfnamefont {A.~O.}\ \bibnamefont
  {Ivanov}}\ and\ \bibinfo {author} {\bibfnamefont {O.~B.}\ \bibnamefont
  {Kuznetsova}},\ }\href@noop {} {\bibfield  {journal} {\bibinfo  {journal}
  {Phys. Rev. E}\ }\textbf {\bibinfo {volume} {64}},\ \bibinfo {pages} {041405}
  (\bibinfo {year} {2001})}\BibitemShut {NoStop}%
\bibitem [{\citenamefont {Shliomis}(1972)}]{shliomis1972}%
  \BibitemOpen
  \bibfield  {author} {\bibinfo {author} {\bibfnamefont {M.~I.}\ \bibnamefont
  {Shliomis}},\ }\href@noop {} {\bibfield  {journal} {\bibinfo  {journal} {Sov.
  Phys. JETP}\ }\textbf {\bibinfo {volume} {34}},\ \bibinfo {pages} {1291}
  (\bibinfo {year} {1972})}\BibitemShut {NoStop}%
\bibitem [{\citenamefont {Rosensweig}(1985{\natexlab{b}})}]{rosensweig1985}%
  \BibitemOpen
  \bibfield  {author} {\bibinfo {author} {\bibfnamefont {R.~E.}\ \bibnamefont
  {Rosensweig}},\ }\href@noop {} {\bibfield  {journal} {\bibinfo  {journal}
  {Journal of Applied Physics}\ }\textbf {\bibinfo {volume} {57}},\ \bibinfo
  {pages} {4259} (\bibinfo {year} {1985}{\natexlab{b}})}\BibitemShut {NoStop}%
\bibitem [{\citenamefont {Zelazo}\ and\ \citenamefont
  {Melcher}(1969)}]{zelazo69}%
  \BibitemOpen
  \bibfield  {author} {\bibinfo {author} {\bibfnamefont {R.~E.}\ \bibnamefont
  {Zelazo}}\ and\ \bibinfo {author} {\bibfnamefont {J.~R.}\ \bibnamefont
  {Melcher}},\ }\href@noop {} {\bibfield  {journal} {\bibinfo  {journal} {J.
  Fluid Mech.}\ }\textbf {\bibinfo {volume} {39}},\ \bibinfo {pages} {1}
  (\bibinfo {year} {1969})}\BibitemShut {NoStop}%
\bibitem [{\citenamefont {Rault}(2000)}]{rault00}%
  \BibitemOpen
  \bibfield  {author} {\bibinfo {author} {\bibfnamefont {J.}~\bibnamefont
  {Rault}},\ }\href@noop {} {\bibfield  {journal} {\bibinfo  {journal} {J.
  Non-Cryst. Solids}\ }\textbf {\bibinfo {volume} {271}},\ \bibinfo {pages}
  {177} (\bibinfo {year} {2000})}\BibitemShut {NoStop}%
\bibitem [{\citenamefont {Tanner}(2000)}]{tanner2000}%
  \BibitemOpen
  \bibfield  {author} {\bibinfo {author} {\bibfnamefont {R.~I.}\ \bibnamefont
  {Tanner}},\ }\href@noop {} {\emph {\bibinfo {title} {Engineering rheology}}}\
  (\bibinfo  {publisher} {Oxford University Press},\ \bibinfo {year}
  {2000})\BibitemShut {NoStop}%
\bibitem [{\citenamefont {W.}(1958)}]{sisko1958}%
  \BibitemOpen
  \bibfield  {author} {\bibinfo {author} {\bibfnamefont {A.~W.}\ \bibnamefont
  {Sisko}},\ }\href@noop {} {\bibfield  {journal} {\bibinfo  {journal} {Ind. Eng.
  Chem.}\ }\textbf {\bibinfo {volume} {50}},\ \bibinfo {pages} {1789} (\bibinfo
  {year} {1958})}\BibitemShut {NoStop}%
\bibitem [{\citenamefont {Shliomis}(1974)}]{shliomis74}%
  \BibitemOpen
  \bibfield  {author} {\bibinfo {author} {\bibfnamefont {M.~I.}\ \bibnamefont
  {Shliomis}},\ }\href@noop {} {\bibfield  {journal} {\bibinfo  {journal} {Usp.
  Fiz. Nauk}\ }\textbf {\bibinfo {volume} {112}},\ \bibinfo {pages} {427}
  (\bibinfo {year} {1974})},\ \bibinfo {note} {[Sov. Phys. Usp. {\bf 17}, 153
  (1974)]}\BibitemShut {NoStop}%
\bibitem [{\citenamefont {Klokkenburg}\ \emph {et~al.}(2006)\citenamefont
  {Klokkenburg}, \citenamefont {Dullens}, \citenamefont {Kegel}, \citenamefont
  {Erne},\ and\ \citenamefont {Philipse}}]{klokkenburg2006}%
  \BibitemOpen
  \bibfield  {author} {\bibinfo {author} {\bibfnamefont {M.}~\bibnamefont
  {Klokkenburg}}, \bibinfo {author} {\bibfnamefont {R.}~\bibnamefont
  {Dullens}}, \bibinfo {author} {\bibfnamefont {W.}~\bibnamefont {Kegel}},
  \bibinfo {author} {\bibfnamefont {B.}~\bibnamefont {Erne}}, \ and\ \bibinfo
  {author} {\bibfnamefont {A.}~\bibnamefont {Philipse}},\ }\href@noop {}
  {\bibfield  {journal} {\bibinfo  {journal} {Phys. Rev. Lett.}\ }\textbf
  {\bibinfo {volume} {96}},\ \bibinfo {pages} {037203 1} (\bibinfo {year}
  {2006})}\BibitemShut {NoStop}%
\bibitem [{\citenamefont {Odenbach}\ and\ \citenamefont
  {St\"ork}(1998)}]{odenbach1998}%
  \BibitemOpen
  \bibfield  {author} {\bibinfo {author} {\bibfnamefont {S.}~\bibnamefont
  {Odenbach}}\ and\ \bibinfo {author} {\bibfnamefont {H.}~\bibnamefont
  {St\"ork}},\ }\href@noop {} {\bibfield  {journal} {\bibinfo  {journal} {J.
  Magn. Magn. Mater.}\ }\textbf {\bibinfo {volume} {183}},\ \bibinfo {pages}
  {188} (\bibinfo {year} {1998})}\BibitemShut {NoStop}%
\bibitem [{\citenamefont {Odenbach}(2009)}]{odenbach2009}%
  \BibitemOpen
  \bibinfo {editor} {\bibfnamefont {S.}~\bibnamefont {Odenbach}},\ ed.,\
  \href@noop {} {\emph {\bibinfo {title} {Colloidal Magnetic Fluids: Basics,
  Development and Applications of Ferrofluids}}},\ \bibinfo {series} {Lect.
  Notes Phys.}, Vol.\ \bibinfo {volume} {763}\ (\bibinfo  {publisher}
  {Springer},\ \bibinfo {address} {Berlin, Heidelberg, New York},\ \bibinfo
  {year} {2009})\BibitemShut {NoStop}%
\bibitem [{\citenamefont {Gollwitzer}\ \emph
  {et~al.}(2007{\natexlab{b}})\citenamefont {Gollwitzer}, \citenamefont
  {Matthies}, \citenamefont {Richter}, \citenamefont {Rehberg},\ and\
  \citenamefont {Tobiska}}]{gollwitzer2007}%
  \BibitemOpen
  \bibfield  {author} {\bibinfo {author} {\bibfnamefont {C.}~\bibnamefont
  {Gollwitzer}}, \bibinfo {author} {\bibfnamefont {G.}~\bibnamefont
  {Matthies}}, \bibinfo {author} {\bibfnamefont {R.}~\bibnamefont {Richter}},
  \bibinfo {author} {\bibfnamefont {I.}~\bibnamefont {Rehberg}}, \ and\
  \bibinfo {author} {\bibfnamefont {L.}~\bibnamefont {Tobiska}},\ }\href@noop
  {} {\bibfield  {journal} {\bibinfo  {journal} {J. Fluid Mech.}\ }\textbf
  {\bibinfo {volume} {571}},\ \bibinfo {pages} {455} (\bibinfo {year}
  {2007}{\natexlab{b}})}\BibitemShut {NoStop}%
\bibitem [{\citenamefont {Cao}\ and\ \citenamefont {Ding}(2014)}]{cao14}%
  \BibitemOpen
  \bibfield  {author} {\bibinfo {author} {\bibfnamefont {Y.}~\bibnamefont
  {Cao}}\ and\ \bibinfo {author} {\bibfnamefont {Z.}~\bibnamefont {Ding}},\
  }\href@noop {} {\bibfield  {journal} {\bibinfo  {journal} {J. Magn. Magn.
  Mater.}\ }\textbf {\bibinfo {volume} {355}},\ \bibinfo {pages} {93} (\bibinfo
  {year} {2014})}\BibitemShut {NoStop}%
\bibitem [{\citenamefont {Gollwitzer}\ \emph {et~al.}(2006)\citenamefont
  {Gollwitzer}, \citenamefont {Rehberg},\ and\ \citenamefont
  {Richter}}]{gollwitzer06}%
  \BibitemOpen
  \bibfield  {author} {\bibinfo {author} {\bibfnamefont {C.}~\bibnamefont
  {Gollwitzer}}, \bibinfo {author} {\bibfnamefont {I.}~\bibnamefont {Rehberg}},
  \ and\ \bibinfo {author} {\bibfnamefont {R.}~\bibnamefont {Richter}},\
  }\href@noop {} {\bibfield  {journal} {\bibinfo  {journal} {J. Phys.: Condens.
  Matter}\ }\textbf {\bibinfo {volume} {18}},\ \bibinfo {pages} {S2643}
  (\bibinfo {year} {2006})}\BibitemShut {NoStop}%
\bibitem [{\citenamefont {Friedrich}\ \emph {et~al.}(2011)\citenamefont
  {Friedrich}, \citenamefont {Lange}, \citenamefont {Rehberg},\ and\
  \citenamefont {Richter}}]{friedrich11}%
  \BibitemOpen
  \bibfield  {author} {\bibinfo {author} {\bibfnamefont {T.}~\bibnamefont
  {Friedrich}}, \bibinfo {author} {\bibfnamefont {A.}~\bibnamefont {Lange}},
  \bibinfo {author} {\bibfnamefont {I.}~\bibnamefont {Rehberg}}, \ and\
  \bibinfo {author} {\bibfnamefont {R.}~\bibnamefont {Richter}},\ }\href@noop
  {} {\bibfield  {journal} {\bibinfo  {journal} {Magnetohydrodynamics}\
  }\textbf {\bibinfo {volume} {47}},\ \bibinfo {pages} {167} (\bibinfo {year}
  {2011})}\BibitemShut {NoStop}%
\bibitem [{\citenamefont {Lavrova}\ \emph {et~al.}(2011)\citenamefont
  {Lavrova}, \citenamefont {Polevikov},\ and\ \citenamefont
  {Tobiska}}]{lavrova10}%
  \BibitemOpen
  \bibfield  {author} {\bibinfo {author} {\bibfnamefont {O.}~\bibnamefont
  {Lavrova}}, \bibinfo {author} {\bibfnamefont {V.}~\bibnamefont {Polevikov}},
  \ and\ \bibinfo {author} {\bibfnamefont {L.}~\bibnamefont {Tobiska}},\
  }\href@noop {} {\bibfield  {journal} {\bibinfo  {journal} {Math. Modell.
  Anal.}\ }\textbf {\bibinfo {volume} {15}},\ \bibinfo {pages} {223} (\bibinfo
  {year} {2011})}\BibitemShut {NoStop}%
\bibitem [{\citenamefont {Reimann}\ \emph {et~al.}(2003)\citenamefont
  {Reimann}, \citenamefont {Richter}, \citenamefont {Rehberg},\ and\
  \citenamefont {Lange}}]{reimann03}%
  \BibitemOpen
  \bibfield  {author} {\bibinfo {author} {\bibfnamefont {B.}~\bibnamefont
  {Reimann}}, \bibinfo {author} {\bibfnamefont {R.}~\bibnamefont {Richter}},
  \bibinfo {author} {\bibfnamefont {I.}~\bibnamefont {Rehberg}}, \ and\
  \bibinfo {author} {\bibfnamefont {A.}~\bibnamefont {Lange}},\ }\href@noop {}
  {\bibfield  {journal} {\bibinfo  {journal} {Phys. Rev. E}\ }\textbf {\bibinfo
  {volume} {68}},\ \bibinfo {pages} {036220} (\bibinfo {year}
  {2003})}\BibitemShut {NoStop}%
\bibitem [{\citenamefont {Lange}(2003)}]{lange_mhd03}%
  \BibitemOpen
  \bibfield  {author} {\bibinfo {author} {\bibfnamefont {A.}~\bibnamefont
  {Lange}},\ }\href@noop {} {\bibfield  {journal} {\bibinfo  {journal}
  {Magnetohydrodynamics}\ }\textbf {\bibinfo {volume} {39}},\ \bibinfo {pages}
  {65} (\bibinfo {year} {2003})}\BibitemShut {NoStop}%
\bibitem [{com()}]{comment1_PRE15}%
  \BibitemOpen
  \href@noop {} {}\bibinfo {note} {Due to the equivalenz of $\hat B$ and $\hat
  M$, the scaling law $\hat k_\mathrm{m}=c_3 \hat B + c_4 \sqrt{\hat B}$ in
  \protect\cite{lange01_growth} is adapted to $\hat k_\mathrm{m}=\tilde c_3
  \hat M + \tilde c_4 \sqrt{\hat M}$.}\BibitemShut {Stop}%
\bibitem [{\citenamefont {Press}\ \emph {et~al.}(2002)\citenamefont {Press},
  \citenamefont {Teukolsky}, \citenamefont {Vetterling},\ and\ \citenamefont
  {Flannery}}]{press2002numerical}%
  \BibitemOpen
  \bibfield  {author} {\bibinfo {author} {\bibfnamefont {H.}~\bibnamefont
  {Press}}, \bibinfo {author} {\bibfnamefont {A.}~\bibnamefont {Teukolsky}},
  \bibinfo {author} {\bibfnamefont {T.}~\bibnamefont {Vetterling}}, \ and\
  \bibinfo {author} {\bibfnamefont {P.}~\bibnamefont {Flannery}},\ }\href@noop
  {} {\emph {\bibinfo {title} {Numerical Recipes in C++. The Art of Computer
  Programming}}}\ (\bibinfo  {publisher} {Cambridge University Press New
  York},\ \bibinfo {year} {2002})\BibitemShut {NoStop}%
\bibitem [{\citenamefont {Ivanov}\ and\ \citenamefont
  {Pshenichnikov}(2010)}]{ivanov10}%
  \BibitemOpen
  \bibfield  {author} {\bibinfo {author} {\bibfnamefont {A.~S.}\ \bibnamefont
  {Ivanov}}\ and\ \bibinfo {author} {\bibfnamefont {A.~F.}\ \bibnamefont
  {Pshenichnikov}},\ }\href@noop {} {\bibfield  {journal} {\bibinfo  {journal}
  {J. Magn. Magn. Mater.}\ }\textbf {\bibinfo {volume} {322}},\ \bibinfo
  {pages} {2575} (\bibinfo {year} {2010})}\BibitemShut {NoStop}%
\bibitem [{\citenamefont {Kityk}\ and\ \citenamefont
  {Wagner}(2006)}]{kityk2006}%
  \BibitemOpen
  \bibfield  {author} {\bibinfo {author} {\bibfnamefont {A.}~\bibnamefont
  {Kityk}}\ and\ \bibinfo {author} {\bibfnamefont {C.}~\bibnamefont {Wagner}},\
  }\href@noop {} {\bibfield  {journal} {\bibinfo  {journal} {Europhys.\,Lett.}\
  }\textbf {\bibinfo {volume} {75}},\ \bibinfo {pages} {441} (\bibinfo {year}
  {2006})}\BibitemShut {NoStop}%
\bibitem [{\citenamefont {Galindo-Rosales}\ \emph {et~al.}(2015)\citenamefont
  {Galindo-Rosales}, \citenamefont {Segovia-Gutiérrez}, \citenamefont {Pinho},
  \citenamefont {Alves},\ and\ \citenamefont {de~Vicente}}]{galindo2015}%
  \BibitemOpen
  \bibfield  {author} {\bibinfo {author} {\bibfnamefont {F.~J.}\ \bibnamefont
  {Galindo-Rosales}}, \bibinfo {author} {\bibfnamefont {J.~P.}\ \bibnamefont
  {Segovia-Gutiérrez}}, \bibinfo {author} {\bibfnamefont {F.~T.}\ \bibnamefont
  {Pinho}}, \bibinfo {author} {\bibfnamefont {M.~A.}\ \bibnamefont {Alves}}, \
  and\ \bibinfo {author} {\bibfnamefont {J.}~\bibnamefont {de~Vicente}},\
  }\href {\doibase http://dx.doi.org/10.1122/1.4902356} {\bibfield  {journal}
  {\bibinfo  {journal} {J. Rheol.}\ }\textbf {\bibinfo {volume} {59}},\
  \bibinfo {pages} {193} (\bibinfo {year} {2015})}\BibitemShut {NoStop}%
\end{thebibliography}

%

\end{document}